\documentclass[aps,prd,reprint,superscriptaddress]{revtex4-1}

\usepackage{graphicx}
\usepackage{amsmath}
\usepackage{amssymb}
\usepackage{units}
\usepackage{color}
\usepackage{hyperref}
\usepackage[caption=false]{subfig}

\newcommand{\icxlong}{\emph{IceCube-Gen2}}
\newcommand{\icx}{\emph{IceCube-Gen2}}

\begin{document}

\title{\icxlong{}: A Vision for the Future of Neutrino Astronomy in Antarctica}

\affiliation{III. Physikalisches Institut, RWTH Aachen University, D-52056 Aachen, Germany}
\affiliation{School of Chemistry \& Physics, University of Adelaide, Adelaide SA, 5005 Australia}
\affiliation{Dept.~of Physics and Astronomy, University of Alaska Anchorage, 3211 Providence Dr., Anchorage, AK 99508, USA}
\affiliation{CTSPS, Clark-Atlanta University, Atlanta, GA 30314, USA}
\affiliation{School of Physics and Center for Relativistic Astrophysics, Georgia Institute of Technology, Atlanta, GA 30332, USA}
\affiliation{Dept.~of Physics, Southern University, Baton Rouge, LA 70813, USA}
\affiliation{Dept.~of Physics, University of California, Berkeley, CA 94720, USA}
\affiliation{Lawrence Berkeley National Laboratory, Berkeley, CA 94720, USA}
\affiliation{Institut f\"ur Physik, Humboldt-Universit\"at zu Berlin, D-12489 Berlin, Germany}
\affiliation{Fakult\"at f\"ur Physik \& Astronomie, Ruhr-Universit\"at Bochum, D-44780 Bochum, Germany}
\affiliation{Physikalisches Institut, Universit\"at Bonn, Nussallee 12, D-53115 Bonn, Germany}
\affiliation{Universit\'e Libre de Bruxelles, Science Faculty CP230, B-1050 Brussels, Belgium}
\affiliation{Vrije Universiteit Brussel, Dienst ELEM, B-1050 Brussels, Belgium}
\affiliation{Dept.~of Physics, Massachusetts Institute of Technology, Cambridge, MA 02139, USA}
\affiliation{Dept.~of Physics, Chiba University, Chiba 263-8522, Japan}
\affiliation{Dept.~of Physics and Astronomy, University of Canterbury, Private Bag 4800, Christchurch, New Zealand}
\affiliation{Dept.~of Physics, University of Maryland, College Park, MD 20742, USA}
\affiliation{Dept.~of Physics and Center for Cosmology and Astro-Particle Physics, Ohio State University, Columbus, OH 43210, USA}
\affiliation{Dept.~of Astronomy, Ohio State University, Columbus, OH 43210, USA}
\affiliation{Niels Bohr Institute, University of Copenhagen, DK-2100 Copenhagen, Denmark}
\affiliation{Dept.~of Physics, TU Dortmund University, D-44221 Dortmund, Germany}
\affiliation{Dept.~of Physics and Astronomy, Michigan State University, East Lansing, MI 48824, USA}
\affiliation{Dept.~of Physics, University of Alberta, Edmonton, Alberta, Canada T6G 2E1}
\affiliation{Erlangen Centre for Astroparticle Physics, Friedrich-Alexander-Universit\"at Erlangen-N\"urnberg, D-91058 Erlangen, Germany}
\affiliation{D\'epartement de physique nucl\'eaire et corpusculaire, Universit\'e de Gen\`eve, CH-1211 Gen\`eve, Switzerland}
\affiliation{Dept.~of Physics and Astronomy, University of Gent, B-9000 Gent, Belgium}
\affiliation{Dept.~of Physics and Astronomy, University of California, Irvine, CA 92697, USA}
\affiliation{Dept.~of Physics and Astronomy, University of Kansas, Lawrence, KS 66045, USA}
\affiliation{School of Physics and Astronomy, Queen Mary University of London, London E1 4NS, United Kingdom}
\affiliation{Dept.~of Astronomy, University of Wisconsin, Madison, WI 53706, USA}
\affiliation{Dept.~of Physics and Wisconsin IceCube Particle Astrophysics Center, University of Wisconsin, Madison, WI 53706, USA}
\affiliation{Institute of Physics, University of Mainz, Staudinger Weg 7, D-55099 Mainz, Germany}
\affiliation{The University of Manchester, Manchester, M13 9PL, UK}
\affiliation{Universit\'e de Mons, 7000 Mons, Belgium}
\affiliation{Technische Universit\"at M\"unchen, D-85748 Garching, Germany}
\affiliation{Columbia Astrophysics and Nevis Laboratories, Columbia University, New York, NY 10027, USA}
\affiliation{Bartol Research Institute and Dept.~of Physics and Astronomy, University of Delaware, Newark, DE 19716, USA}
\affiliation{Dept.~of Physics, Yale University, New Haven, CT 06520, USA}
\affiliation{Dept.~of Physics, University of Notre Dame du Lac, 225 Nieuwland Science Hall, Notre Dame, IN 46556-5670, USA}
\affiliation{Dept.~of Physics, University of Oxford, 1 Keble Road, Oxford OX1 3NP, UK}
\affiliation{Dept.~of Physics, Drexel University, 3141 Chestnut Street, Philadelphia, PA 19104, USA}
\affiliation{Dept.~of Physics, South Dakota School of Mines and Technology, Rapid City, SD 57701, USA}
\affiliation{Dept.~of Physics, University of Wisconsin, River Falls, WI 54022, USA}
\affiliation{Oskar Klein Centre and Dept.~of Physics, Stockholm University, SE-10691 Stockholm, Sweden}
\affiliation{Dept.~of Physics and Astronomy, Stony Brook University, Stony Brook, NY 11794-3800, USA}
\affiliation{Dept.~of Physics, Sungkyunkwan University, Suwon 440-746, Korea}
\affiliation{Earthquake Research Institute, University of Tokyo, Bunkyo, Tokyo 113-0032, Japan}
\affiliation{Dept.~of Physics, University of Toronto, Toronto, Ontario, Canada, M5S 1A7}
\affiliation{Dept.~of Physics and Astronomy, University of Alabama, Tuscaloosa, AL 35487, USA}
\affiliation{Dept.~of Astronomy and Astrophysics, Pennsylvania State University, University Park, PA 16802, USA}
\affiliation{Dept.~of Physics, Pennsylvania State University, University Park, PA 16802, USA}
\affiliation{Dept.~of Physics and Astronomy, Uppsala University, Box 516, S-75120 Uppsala, Sweden}
\affiliation{Dept.~of Physics, University of Wuppertal, D-42119 Wuppertal, Germany}
\affiliation{DESY, D-15735 Zeuthen, Germany}

\author{M.~G.~Aartsen}
\affiliation{School of Chemistry \& Physics, University of Adelaide, Adelaide SA, 5005 Australia}
\author{M.~Ackermann}
\affiliation{DESY, D-15735 Zeuthen, Germany}
\author{J.~Adams}
\affiliation{Dept.~of Physics and Astronomy, University of Canterbury, Private Bag 4800, Christchurch, New Zealand}
\author{J.~A.~Aguilar}
\affiliation{Universit\'e Libre de Bruxelles, Science Faculty CP230, B-1050 Brussels, Belgium}
\author{M.~Ahlers}
\affiliation{Dept.~of Physics and Wisconsin IceCube Particle Astrophysics Center, University of Wisconsin, Madison, WI 53706, USA}
\author{M.~Ahrens}
\affiliation{Oskar Klein Centre and Dept.~of Physics, Stockholm University, SE-10691 Stockholm, Sweden}
\author{D.~Altmann}
\affiliation{Erlangen Centre for Astroparticle Physics, Friedrich-Alexander-Universit\"at Erlangen-N\"urnberg, D-91058 Erlangen, Germany}
\author{T.~Anderson}
\affiliation{Dept.~of Physics, Pennsylvania State University, University Park, PA 16802, USA}
\author{G.~Anton}
\affiliation{Erlangen Centre for Astroparticle Physics, Friedrich-Alexander-Universit\"at Erlangen-N\"urnberg, D-91058 Erlangen, Germany}
\author{C.~Arguelles}
\affiliation{Dept.~of Physics and Wisconsin IceCube Particle Astrophysics Center, University of Wisconsin, Madison, WI 53706, USA}
% ...

\collaboration{IceCube-Gen2 Collaboration}
\noaffiliation

\begin{abstract}
The recent observation by the IceCube neutrino observatory of an astrophysical flux of neutrinos represents the ``first light'' in the nascent field of
neutrino astronomy.  The observed diffuse neutrino flux seems to suggest a much larger level of hadronic activity in the non-thermal universe than previously thought and suggests a rich discovery potential for a larger neutrino observatory.  
This document presents a vision for an substantial expansion of the current IceCube detector, \icxlong{}, including the aim of instrumenting a $10\,\mathrm{km}^3$ volume
of clear glacial ice at the South Pole to deliver substantial increases in the astrophysical neutrino sample for all flavors.  A detector of this size
would have a rich physics program with the goal to resolve the sources of these astrophysical neutrinos, discover GZK neutrinos,
and be a leading observatory in future multi-messenger astronomy programs.  
\end{abstract}

\maketitle

\section*{Executive Summary}

Developments in neutrino astronomy have been driven by the search for the sources of cosmic rays,
leading, at an early stage, to the concept of a cubic kilometer neutrino detector. Four decades later, IceCube
has discovered a flux of high-energy neutrinos of cosmic origin~\cite{Aartsen:2013jdh, Aartsen:2014gkd}. 
The observed neutrino flux implies that a significant fraction of the energy in the non-thermal universe, powered by the gravitational energy of compact objects from neutron stars to supermassive black holes, is generated in hadronic accelerators. High-energy neutrinos therefore hold the discovery potential to either reveal new sources or provide new insight into the energy generation of known sources.

The observed spectrum of neutrinos, resulting from general agreement among a sequence of independent analyses of multiple years of IceCube data, has revealed approximately 100 astrophysical neutrino events. The ability of IceCube to be an efficient tool for neutrino astronomy over the next decade is limited by the modest numbers of cosmic neutrinos measured, even in a cubic kilometer array. In this paper we present a vision for the next-generation IceCube neutrino observatory, at the heart of which is an expanded array of light-sensing modules that instrument a $10\,\mathrm{km}^3$ volume for detection of high-energy neutrinos.  With its unprecedented sensitivity and improved angular resolution, this instrument will explore extreme energies (PeV-scale) and will collect high-statistics samples of astrophysical neutrinos of all flavors, enabling detailed spectral studies, significant point source detections and new discoveries. 
The large gain in event rate is made possible by the unique optical properties of the Antarctic glacier revealed by the construction and operation of IceCube. Extremely long photon absorption lengths in the deep ice means the spacing between strings of light sensors may exceed $250\,\mathrm{m}$, enabling the instrumented volume to grow rapidly while the cost for the high-energy array remains comparable to that of the current IceCube detector. By roughly doubling the instrumentation already deployed, a telescope with an instrumented volume of
$10\,\mathrm{km}^3$ is achievable and will yield a significant increase in astrophysical neutrino detection rates. The instrument will provide an unprecedented view of the high-energy universe, taking neutrino astronomy to new levels of discovery with the potential to resolve the question of the origin of the cosmic neutrinos recently discovered~\cite{Aartsen:2013jdh, Aartsen:2014gkd}. 

By delivering a significantly larger sample of high-energy neutrinos with improved angular resolution and measurement of the energy, a detailed understanding of the source distribution, spectrum and flavor composition of the astrophysical neutrinos is within reach. This sample will reveal an unobstructed view of the universe at $>$PeV energy, previously unexplored wavelengths where most of the universe is opaque to high-energy photons. The operation of a next-generation IceCube detector in coincidence with the next generations of optical-to-gamma-ray telescopes and gravitational wave detectors will present novel opportunities for multi-messenger astronomy and multi-wavelength follow-up campaigns to obtain a  complete picture of astrophysical sources.

Because of its sheer size, the high-energy array has the potential to deliver significant samples of EeV-energy GZK neutrinos, of anti-electron neutrinos produced via the Glashow resonance~\cite{Glashow:1960zz}, and of PeV tau neutrinos, where both particle showers associated with the production and decay of the tau are observed. GZK neutrinos produced in interactions of extragalactic cosmic rays with microwave photons are within reach of the instrument provided a fraction (at least at the 10\% level) of the extragalactic cosmic rays are protons. Their observation will complement PeV neutrino astronomy and may yield a measurement of the neutrino cross-section at center-of-mass energies of $100\,\mathrm{TeV}$, testing electroweak physics at energies beyond the reach of terrestrial accelerators.

Neutrino astronomy will be one one of many topics in the rich science program of a next-generation neutrino observatory. In addition to studying the properties of cosmic rays and searching for  signatures of beyond-the-standard-model neutrino physics, this world-class, multi-purpose detector remains a discovery instrument for new physics and astrophysics. For instance, the observation of neutrinos from a supernova in our galactic neighborhood, in coincidence with astronomical and gravitational wave instruments, would be the astronomical event of the century, providing an unprecedented wealth of information about this key astrophysical process. 

The proposed \icxlong{} high-energy array is envisioned to be the major element of a planned large-scale enhancement to the IceCube facility at the South Pole station.  Members of the \icx{} Collaboration, which is now being formed, are working to develop proposals in the US and elsewhere that will include, besides this next generation IceCube high-energy detector, the PINGU sub-array~\cite{Aartsen:2014oha} that targets precision measurements of the atmospheric oscillation parameters and the determination of the neutrino mass hierarchy. The facility's reach may further be enhanced by exploiting the air-shower measurement and vetoing capabilities of an extended surface array and a radio array to achieve improved sensitivity to neutrinos in the $10^{16}$-$10^{20}\,\mathrm{eV}$ energy range, including GZK neutrinos.

While details of the design of the \icx{} high-energy array, such as the inter-string separations and deployment geometries, remain to be finalized, key elements of its baseline design are robust.  The hot water drilling systems that deploys instrumentation deep into the Antarctic glacier and the digital optical module that records the light radiated by secondary particles produced in neutrino interactions are the key elements for the construction of \icx{}.  Based upon the highly successful designs of the IceCube project, minimal modifications will target improvements focused on modernization, efficiency, and cost savings. These robust baseline designs allow for construction of \icx{} with exceptionally low levels of cost and schedule risk while still exploring new concepts for light sensors in parallel.  Further, due to its digital architecture, the next-generation facility can be operated jointly with the IceCube detector without a significant increase in operational costs. 

The path forward is clear. A complete preliminary design for the \icx{} high-energy array that combines the robust systems for drilling and detector instrumentation with an optimized deployment arrangement that maximizes sensitivity to these newly found astrophysical neutrinos will evolve in the near future. Once in operation, the \icx{} high-energy array, as part of the larger \icx{} facility at the South Pole, will truly be the flagship experiment of the emerging field of neutrino astronomy.

\section{Introduction\label{sec:intro}}

\subsection{IceCube: the First Kilometer-Scale Neutrino Detector}

High-energy neutrinos have a unique potential to probe the extreme universe. Neutrinos reach us from the edge of the universe without absorption or deflection by magnetic fields. They can escape unscathed from the inner neighborhood of black holes and from the accelerators where cosmic rays are born. Their weak interactions make neutrinos very difficult to detect.  By the 1970s, it had been understood \cite{Roberts:1992re} that a kilometer-scale detector was needed to observe the GZK neutrinos produced in the interactions of cosmic rays with background microwave photons \cite{Beresinsky:1969qj}. Today's estimates of neutrino fluxes from potential cosmic ray accelerators such as galactic supernova
remnants, active galactic nuclei (AGN), and gamma-ray bursts (GRB) point to the same size requirement \cite{Gaisser:1994yf, Learned:2000sw, Halzen:2002pg, Becker:2007sv, Katz:2011ke, Halzen:2013bta, 2014ARNPS..64..101G}.
Building such a neutrino telescope has been a daunting technical challenge, focusing on instrumentation of large natural volumes of water or ice to observe the Cherenkov light emitted by the secondary  particles produced when neutrinos interact with nuclei inside or near the detector \cite{Zheleznykh:2006fh, Markov1961385}. 

Early efforts focused on deep-water-based detectors include DUMAND\cite{Babson:1989yy}, Lake Baikal\cite{Balkanov:2000cf},  and ANTARES\cite{Aggouras:2005bg,Aguilar:2006rm,Migneco:2008zz}, which have paved the way toward the proposed construction of KM3NeT\cite{Margiotta:2014eaa} in the Mediterranean sea
and GVD\cite{Avrorin:2013uyc} in Lake Baikal, both with complementary fields of view to that of IceCube.   The deep ice of the Antarctic glacier is host to the first kilometer-scale neutrino observatory.
IceCube\cite{ICPDD2001,Ahrens:2003ix}, completed and in full operation since 2010, builds upon the pioneering work of the Antarctic Muon and Neutrino Detector Array (AMANDA)\cite{Halzen:1998bp} and has begun to probe signals from astrophysical neutrinos.

\begin{figure}
\includegraphics[width=\linewidth]{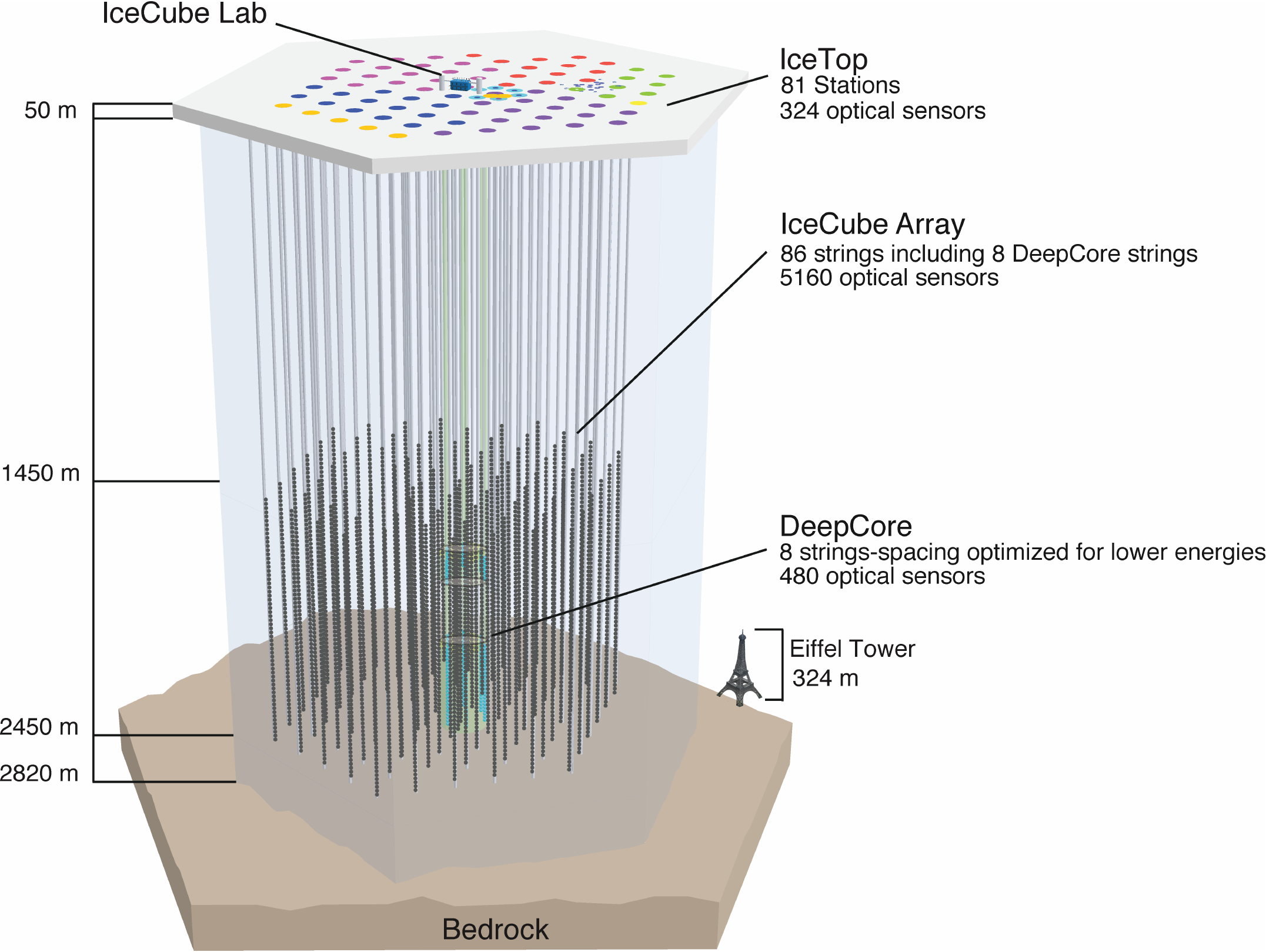}
\caption{Schematic of the IceCube detector.}
\label{deepcore}
\label{fig:IceCube}
\end{figure}

The IceCube neutrino detector (Fig.~\ref{deepcore}) consists of 86 strings, each instrumented with 60 ten-inch photomultipliers spaced 17\,m apart over a total length of one kilometer. The deepest modules are located at a depth of 2.45\,km so that the instrument is shielded from the large background of cosmic rays at the surface by approximately 1.5\,km of ice. Strings are arranged at apexes of equilateral triangles that are 125\,m on a side. The instrumented detector volume is a cubic kilometer of dark and highly transparent \cite{Aartsen:2013rt} Antarctic ice.

Each digital optical module (DOM) consists of a glass sphere containing the photomultiplier and electronics that independently digitize the signals locally using an onboard computer. The digitized signals are given a global time stamp accurate to better than 3\,ns and are subsequently transmitted to the surface. Processors at the surface continuously collect the time-stamped signals from the optical modules, and trigger events based on coincident signals seen in several DOMs. The depth of the detector and its projected area determine the trigger rate of approximately 2.7 kHz for penetrating muons produced by interactions of cosmic
rays in the atmosphere above, outnumbering neutrinos by one per million at TeV energies.  The neutrino rate is dominated by neutrinos produced in the Earth's atmosphere.  The first challenge is to select a sufficiently pure sample of neutrinos, the second is to identify the small fraction that are astrophysical in origin.

Event characteristics such as location, direction and energy are reconstructed using the arrival times and intensity of photons recorded in the DOMs reporting signals in an event. The number of Cherenkov photons produced per unit of path length of a charged particle and their distribution in wavelength are well-known quantities~\cite{Beringer:1900zz}.  A detailed understanding of the propagation of the photons in the ice~\cite{Aartsen:2013rt} is required to relate light generated by relativistic particles to light observed in the DOMs. Good reconstruction of tracks in ice has been achieved and is still being improved~\cite{Ahrens:2003fg}. For typical kilometer-length muon tracks, the angular resolution is better than $0.4^\circ$. The photon timing patterns in each DOM provide the information to reconstruct the direction of the secondary electron or tau in shower events with an angular resolution of $\sim15^\circ$~\cite{Aartsen:2013vja}.  

Neutrino events are broadly classified in two groups by their observed Cherenkov light patterns, muon tracks and particle showers (cascades).  Muon tracks are produced by charged current interactions of muon neutrinos while cascades are produced by charged current interactions of electron and tau neutrinos
as well as by neutral current interactions of all flavors. The range for $\nu_\mu$-induced muons is on the order of kilometers, while the characteristic length scale of electromagnetic showers is only tens of meters. IceCube collects $\nu_\mu$-induced muons filtered by the Earth and entering the detector from below. They overwhelmingly originate from the decay of mesons produced in cosmic ray interactions in the Earth's atmosphere.  Atmospheric neutrinos are collected at a rate of $\sim300$  per day, have a mean energy of $\sim1$ TeV and form an irreducible background in searches for astrophysical neutrinos.  At energies in excess of $\sim100$ TeV, the flux of atmospheric neutrinos is small, and events with higher energy are likely of astrophysical origin.

A complementary way to classify neutrino events is to distinguish events that start inside the detector from those in which the neutrino interacts outside, which can be done for neutrino events originating from the entire sky.

\subsection{Neutrinos Associated with Cosmic Ray Accelerators}

Cosmic accelerators are known to produce particles with energies in excess of $10^{20}$\,eV ($100$\,EeV), yet we still do not know where or how they are accelerated~\cite{Sommers:2008ji}.  The bulk of the cosmic rays are galactic in origin, but any association with our Galaxy vanishes at EeV energy where the gyroradius of a proton in the galactic magnetic field exceeds the size of the Galaxy. The cosmic-ray spectrum exhibits a rich structure above an energy of $\sim 0.1$\,EeV, but where the transition to extragalactic cosmic rays occurs remains an open question.  Speculations on the origin of cosmic rays generally concur that the power of the accelerator is supplied by the gravitational energy of a collapsed object. The energy and luminosity required to explain the measured fluxes are immense, and constrain the sources to objects such as supernova remnants or pulsars in our Galaxy, and gamma-ray bursts, starbursts, clusters of galaxies and active galaxies throughout the universe. 

\begin{figure}
\includegraphics[width=\linewidth]{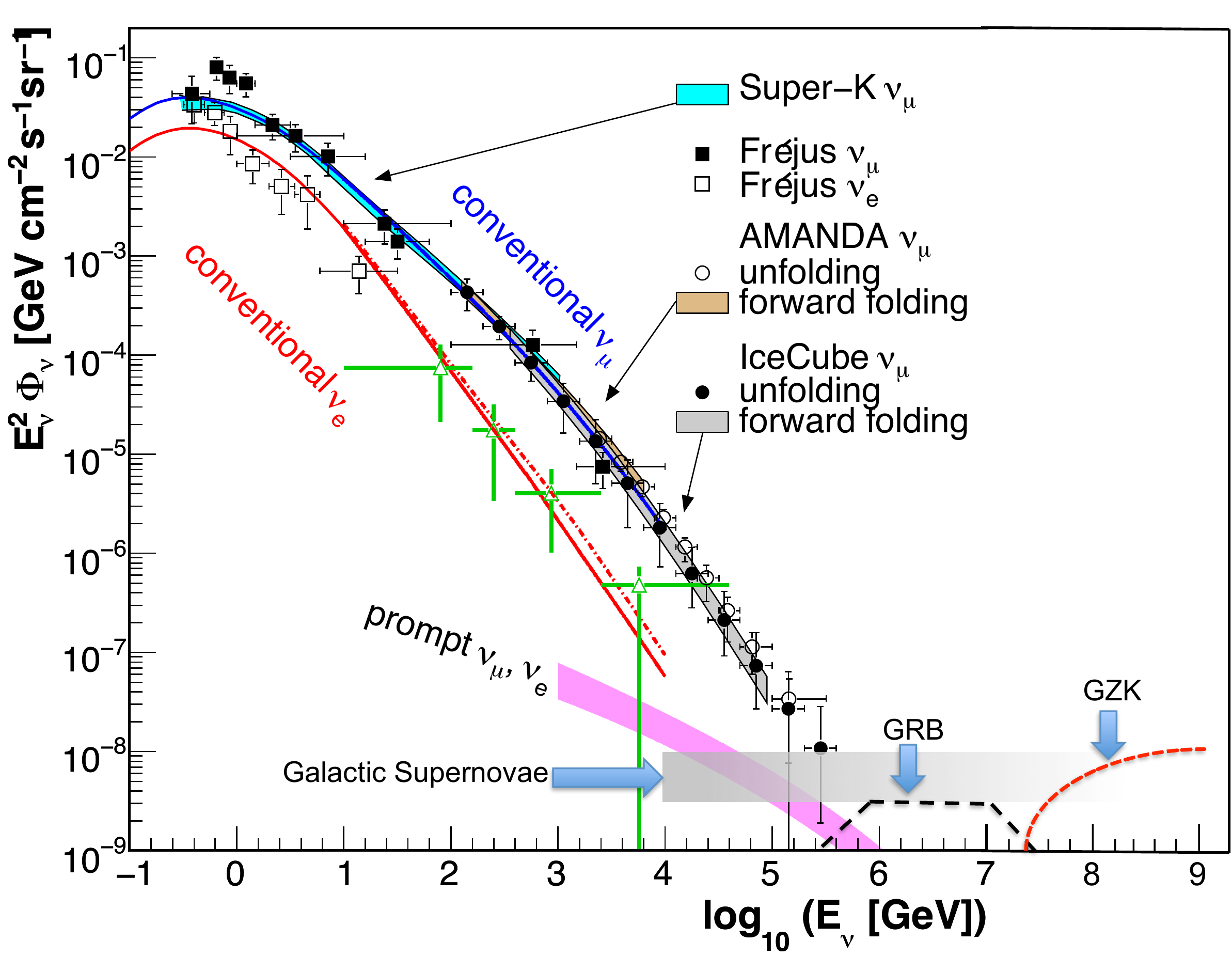}
\caption{Anticipated astrophysical neutrino fluxes produced by supernova remnants and GRBs exceed the atmospheric neutrino flux in
IceCube above 100\, TeV because of their relatively hard $E^{-2}$ spectrum.  Also shown is a sample calculation of the GZK neutrino flux. The atmospheric electron-neutrino
spectrum (green open triangles) is from \cite{Aartsen:2013rt}. The conventional $\nu_e$ (red line) and $\nu_\mu$ (blue line)
from Honda, $\nu_e$ (red dotted line) from Bartol and charm-induced neutrinos (magenta band) \cite{Enberg:2008te} are shown.
Previous measurements from Super-K \cite{GonzalezGarcia:2006ay}, Frejus \cite{Daum:1994bf}, AMANDA \cite{Abbasi:2009nfa,Abbasi:2010qv}
and IceCube \cite{Abbasi:2010ie,Abbasi:2011jx} are also shown. Details about the theoretical estimates shown can be found in Ref.~\cite{Halzen:2013bta}.}
\label{discovery}
\end{figure}

Neutrinos are naturally produced in association with cosmic rays.  Cosmic rays accelerated in regions of high magnetic fields near black holes or neutron stars inevitably interact with radiation and matter surrounding them. In particle physics language, cosmic-ray accelerators and the surrounding material form beam dumps. Estimated neutrino fluxes from several candidate sources are shown in Fig. \ref{discovery}, along with the irreducible atmospheric background as measured prior to the discovery of the astrophysical neutrino flux.   Neutrinos from cosmic-ray accelerators dominate the steeply falling atmospheric neutrino flux above an energy of $\sim100$\,TeV. Assuming an $E^{-2}$ energy spectrum, the anticipated level of astrophysical events observed in a cubic-kilometer neutrino detector is $10\sim100$ per year.   A more detailed description of the theoretical estimates can be found in reference \cite{Halzen:2013bta}.

\subsection{Discovery of Astrophysical Neutrinos}

GZK neutrinos were the target of a dedicated search using IceCube data collected between May 2010 and May 2012. Two events were found \cite{Aartsen:2013bka}. However, their energies were in the PeV range: 1,040 TeV and 1,140 TeV, rather than the EeV energy associated with GZK neutrinos. The events are particle showers initiated by neutrinos interacting inside the instrumented detector volume. Their light pool of roughly one hundred thousand photoelectrons extends over more than 500 meters. With no evidence of an outgoing muon track, they are likely initiated by electron or tau neutrinos.

Inspired by the observation of these events, a filter was designed that exclusively identifies neutrinos that interact inside the detector. It divides the instrumented volume of ice into an outer veto shield and a
420 megaton inner fiducial volume. The separation between veto and signal regions was optimized to reduce the background of atmospheric muons and neutrinos to a handful of events per year while essentially keeping all contained neutrino interactions.   By focusing on neutrinos interacting inside the instrumented volume of ice, the detector functions as a total absorption calorimeter measuring energy
with a 10--15\% resolution. Additionally, neutrinos from all directions in the sky can be identified, including both muon tracks produced in $\nu_\mu$ charged-current interactions and secondary showers produced by neutrinos of all flavors.

Analyzing the data covering the same 2 year period as the GZK neutrino search, 28 candidate neutrino events were identified with in-detector deposited energies between 30 and 1140 TeV, including the two events already found in the search for GZK neutrinos.  Of these, 21 are showers whose energies are measured to better than 15\% but whose directions are determined to 10-15 degrees only. Predominantly originating in the southern hemisphere, none show evidence for an accompanying muon track. If these neutrinos were produced in atmospheric air showers, they would likely be accompanied by muons produced in the parent air shower from which they originate.  For 1~PeV down-going atmospheric neutrinos, where the expectation before the veto is a few events per year, this atmospheric self-vetoing would remove more than 99.9\% of them by observing an accompanying muon in the deep detector in coincidence with the neutrino\cite{Schonert:2008is,Gaisser:2014bja}.

The remaining seven events contain muon tracks, which do provide sub-degree angular resolution, but a precise energy estimate is challenging as a fraction of the energy is carried away by the exiting muon track. Furthermore, with the present statistics, these are difficult to separate clearly from the atmospheric muon background.

Two additional years of data have been taken with the completed detector, and the first of these has been 
analyzed \cite{Aartsen:2014gkd}. The three-year data set, with an exposure of 988 days, contains a total of 36 neutrino candidate events with deposited energies ranging from 30 TeV to 2000 TeV. The 2000 TeV event is the highest energy neutrino ever observed. In combining the three years of data, a purely atmospheric explanation of the data can be excluded at $5.7\sigma$. Interestingly, the statistical significance is limited by the large errors on a possible atmospheric neutrino background from the production and prompt leptonic decay of charmed particles in air showers (see Fig.~\ref{discovery}). There is no direct evidence of such a ``prompt'' flux in any past experiment.

\begin{figure}
\includegraphics[width=\linewidth]{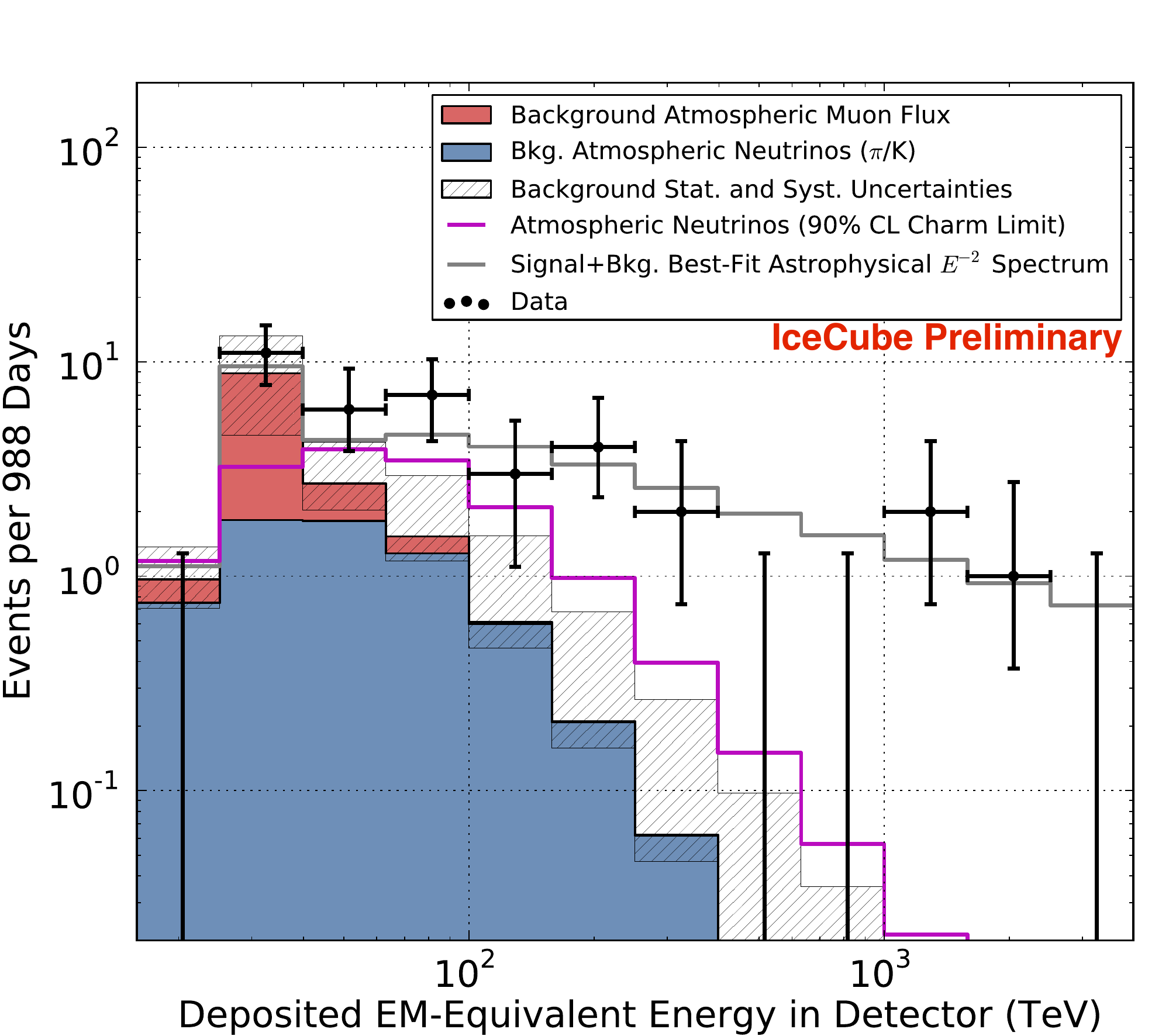}
\caption{Deposited energies of events observed in 3 years of data with predictions. The hashed region shows 
uncertainties on the sum of all backgrounds. Muons (red) are computed from simulation to overcome statistical 
limitations in our background measurement and scaled to match the total measured background rate. Atmospheric 
neutrinos and uncertainties thereon are derived from previous measurements of both the $\pi, K$ and charm components 
of the atmospheric spectrum \cite{Aartsen:2013vca}. A gap larger than the one between 400 and 1000\,TeV appears in 
43\% of realizations of the best-fit continuous spectrum. }
\label{hese_energy}
\end{figure}

Assuming an $E^{-2}$ power-law spectrum, the best fit for 3 years of data to a superposition of astrophysical neutrinos on the atmospheric backgrounds yields an astrophysical neutrino flux of
$E_\nu^2 \frac{dN}{dE_\nu}=2.9\times10^{-8}\,\rm GeV\,cm^{-2}\,s^{-1}\,sr^{-1}$
for the sum of the three neutrino flavors. As will be discussed, this is the level of flux anticipated for neutrinos accompanying the observed cosmic rays \cite{Waxman:1998yy}. Additionally, the energy and zenith angle dependence observed is consistent with what is expected for a flux of neutrinos of astrophysical origin (Fig. \ref{hese_energy}). The flavor composition of the flux is, after corrections for the acceptances of the detector to the different flavors, consistent with 1:1:1 as anticipated for a flux originating in astrophysical sources.

The fourth year of data yielded 17 events that are currently being analyzed.

\begin{figure}
\includegraphics[width=\linewidth]{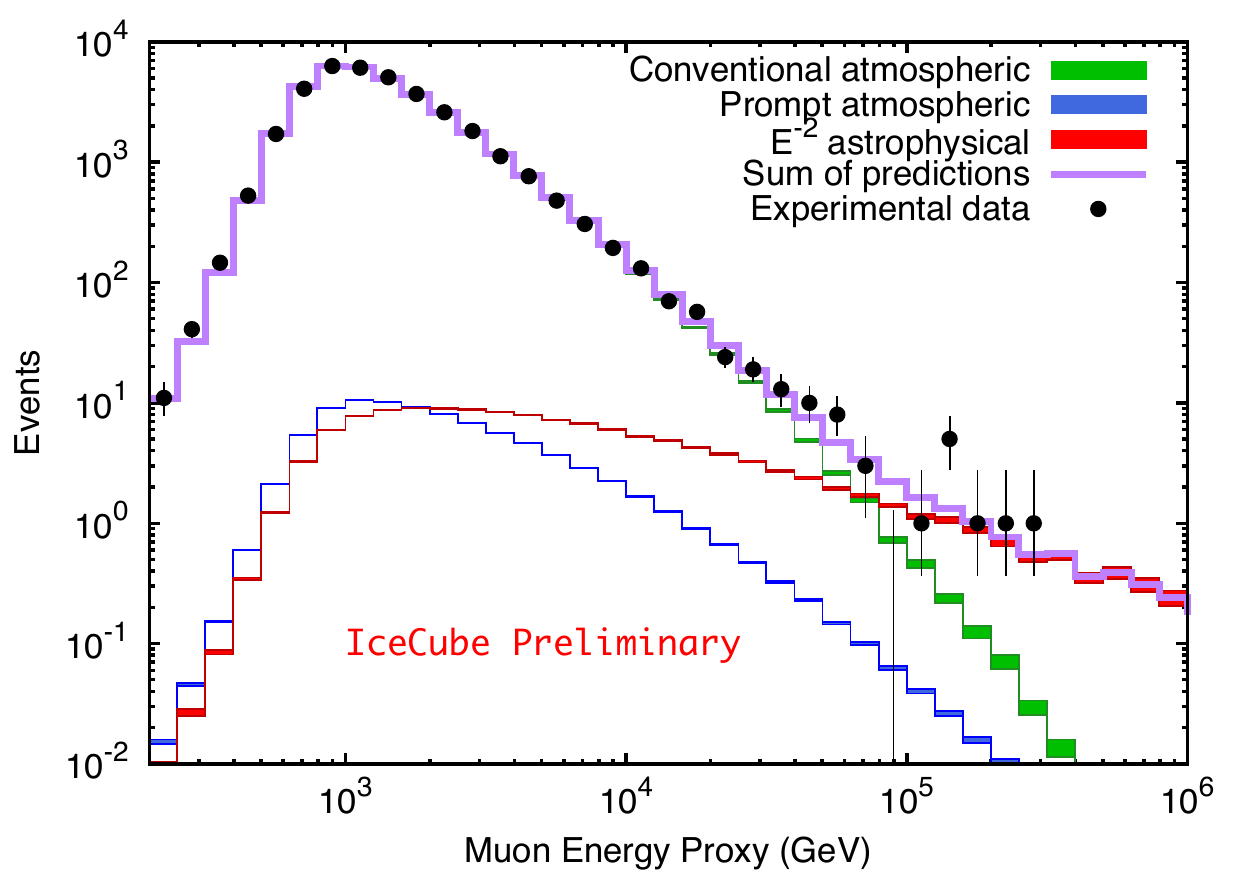}
\caption{Spectrum of secondary muons initiated by muon neutrinos that have traversed the Earth, {\it i.e.}, with zenith 
angle less than $5^\circ$ above the horizon, as a function of the energy deposited inside the detector, used here as a proxy for the muon energy. The 
highest energy muons are, on average, initiated by PeV neutrinos. }
\label{diffusenumu}
\end{figure}

A completely independent analysis of the spectrum of muon neutrinos passing through the Earth has confirmed the existence of the astrophysical component first observed in neutrino events interacting inside the detector.   This search focuses on upward-going muons originating from $\nu_\mu$ charged-current interactions passing through the instrumented volume. Analyzing the same two years of data used for the original starting-event analysis, an excess of high-energy muons tracks is found, rejecting a purely atmospheric neutrino explanation at $3.7\sigma$ \cite{weaver_APS_talk}. The observed spectrum is consistent with the one obtained in the starting event analysis.  Shown in Fig.~\ref{diffusenumu} is the muon neutrino event rate as a function of an energy proxy closely related to the energy deposited by muons inside the detector.  This deposited energy reflects only a fraction of the energy of the neutrino that initiated the events; for instance, the highest energies in Fig.~\ref{diffusenumu} correspond, on average, to parent neutrino energies of $\sim 1\,\mathrm{PeV}$. The best fit to the spectrum including conventional, charm and astrophysical components with unconstrained normalizations is shown in the figure. (Note that the charm component shown is not constrained well by the data in this sample and just represents a fluctuation in the fit compatible with previous upper limits).

This analysis already yields a sample of ten well-reconstructed muon neutrino events in the energy range where the astrophysical flux dominates. In the future it is likely to yield the most powerful data set for searching for point sources. 

A recent analysis of starting events in IceCube~\cite{Aartsen:2014muf} further lowered the energy threshold to about $10\,\mathrm{TeV}$. A fit to the resulting data,  assuming a single unbroken power law, finds a softer spectrum than the $E^{-2}$ benchmark ($E^{-2.46}$), and already mildly excludes a spectral index of 2. The result is also consistent with more complicated spectral shapes and  the best-fit spectral index range of the original contained 3 year analysis of 2.0 -- 2.6.~\cite{Aartsen:2014gkd}

\subsection{Impact of the Discovery of Astrophysical Neutrinos}

\begin{figure}\centering
\includegraphics[width=\linewidth]{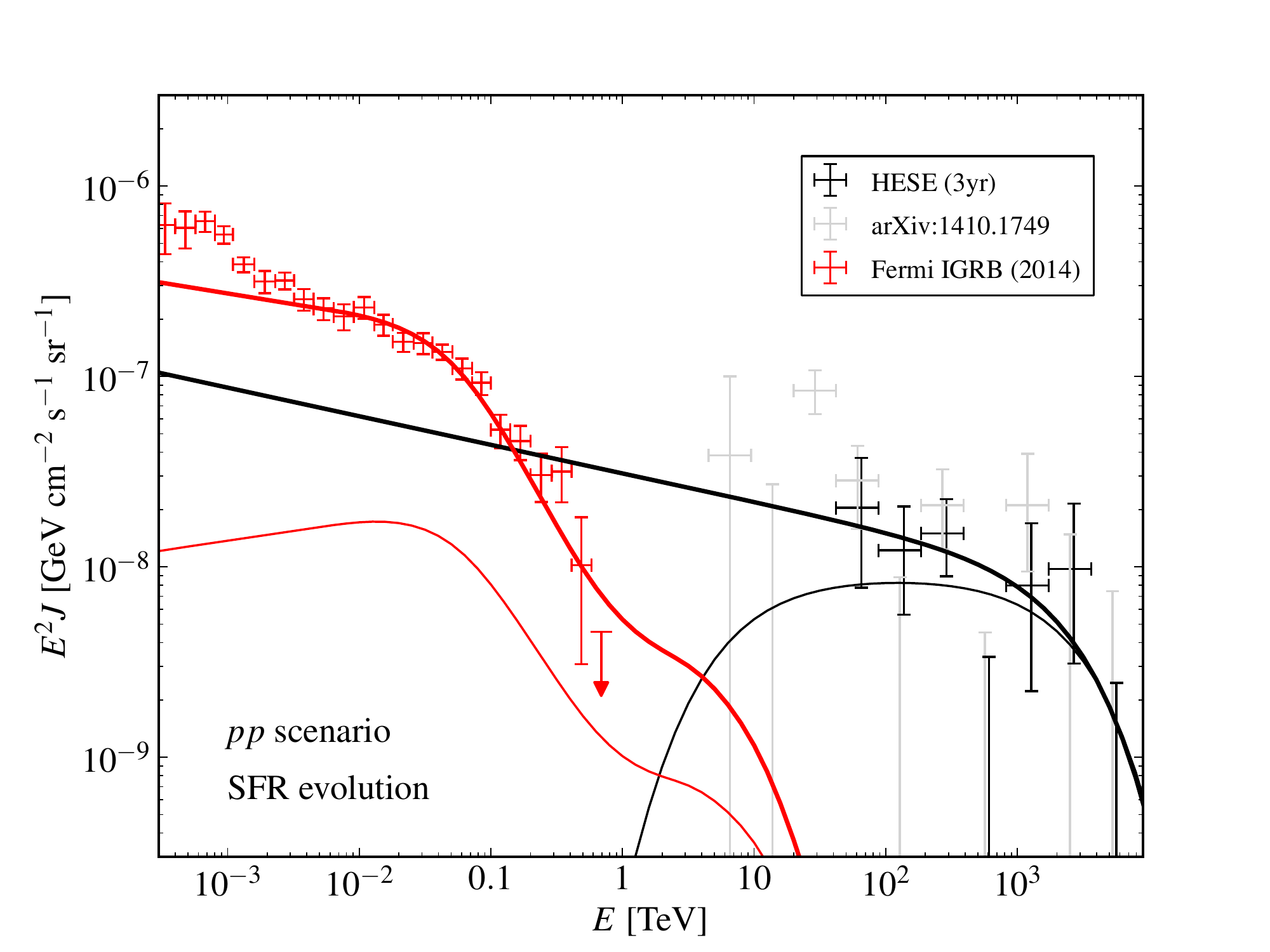}
\caption{Joint fit to the highest energy extragalactic photon flux (red) observed by Fermi and the astrophysical neutrino flux (black) observed by IceCube. The fit assumes that the decay products of neutral and charged pions from $pp$ interactions are responsible for the non-thermal emission in the Universe~\cite{Murase:2013rfa}.  The thin lines represent an attempt to minimize the contribution of the pionic gamma ray flux to the Fermi observations. It assumes an injected flux of $E^{-2}$ with exponential cutoff at low and high energy. The black data points are measured by the IceCube 3-year ``High-Energy Starting Event'' (``HESE'') analysis~\cite{Aartsen:2014gkd}, the gray data points are from an IceCube analysis lowering the energy threshold for events starting in the detector even further~\cite{Aartsen:2014muf}.}
\label{Fermi}
\end{figure}

The most immediate impact of the discovery of astrophysical neutrinos is that the flux level observed is exceptionally high by astronomical standards. 
The magnitude of the observed flux is at a level of the Waxman-Bahcall bound~\cite{Waxman:1998yy} which applies to neutrino production in sources that are also responsible for UHECRs. A saturation of this bound can only be achieved with sources where accelerator and target are essentially integrated, {\it i.e.}~cosmic ray reservoirs~\cite{Katz:2013ooa} such as starburst galaxies~\cite{Loeb:2006tw,Tamborra:2014xia}, clusters of galaxies~\cite{Berezinsky:1996wx,Murase:2008yt,Zandanel:2014pva} or by acceleration near the cores of active galaxies~\cite{Stecker:2013fxa,Kalashev:2014vya}. Such cosmic accelerators produce equal numbers of neutral, positive and negatively charged pions in the proton-proton beam dump. 
The neutral pions accompanying the charged parents of the neutrinos observed by IceCube decay into PeV photons that are only observed indirectly after propagation in the extragalactic background light. Losing energy, they cascade down to energies below 1 TeV observed by the $\gamma$-ray satellite Fermi~\cite{Ackermann:2014usa}.  The relative magnitudes of the diffuse gamma ray flux detected by Fermi and the high energy neutrino flux detected by IceCube imply that most of the energy in the non-thermal universe is produced in hadronic accelerators~\cite{Murase:2013rfa}.  In fact, the assumption that the PeV photons accompanying IceCube neutrinos are the source of {\it all} high energy photons is consistent with the Fermi measurement of the extragalactic flux; see Fig.~\ref{Fermi}.  It is possible to escape this conclusion by assuming that the neutrinos are produced on a photon target by a beam of protons with a relatively flat spectrum. This assumption provides a very poor fit to the neutrino spectrum at lower energies, and even then a hadronic contribution to the high energy diffuse photon flux at the 10\% level is still required.

\begin{figure}
\includegraphics[width=\linewidth]{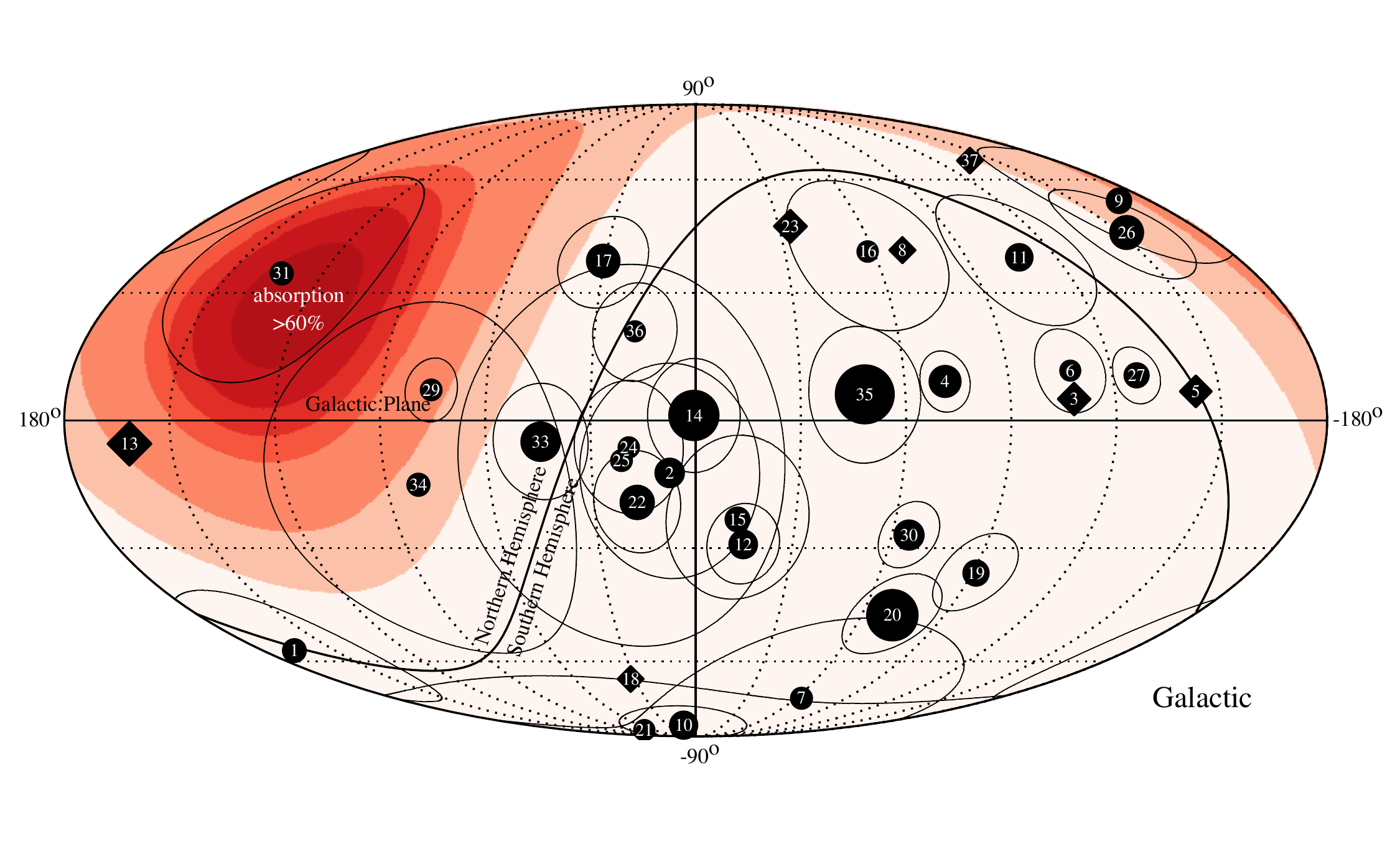}
\caption{Arrival directions of events of the 3yr HESE sample in galactic coordinates. Shower-like events
are shown with filled circles and those containing muon tracks with diamond. The size of the symbols
indicate the deposited energy of the events in the range of 30~TeV to 2~PeV. The thin circles around
cascade events indicate the angular reconstruction uncertainty. The red-shaded regions show 10\%
quantiles of neutrino Earth absorption at 30~TeV and increasing with neutrino energy.
Note, that the track-like event \#28 have been omitted following the discussion
in Ref.~\cite{Aartsen:2014gkd}.}\label{hesemap}
\end{figure}

Where do the observed astrophysical neutrinos originate? Figure~\ref{hesemap} shows the arrival direction of the 3yr ``High-Energy Starting Event'' (HESE) sample in galactic coordinates for cascade events (filled circles) and track events (diamonds).  Anisotropy studies of the arrival direction of HESE events are done by comparing the observed map to  background maps that are obtained by right ascension scrambling of events.  No significant local excess was found. Repeating the analysis for showers only, a hot spot appears close to the galactic center. After correcting for trials, the probability of such a cluster to arise from background alone is 7.2\%. With the present statistics, the events appear to be consistent with an isotropic diffuse flux of neutrinos equally distributed between the three flavors. For instance, a single point source with a $E^{-2}$ spectrum cannot be the origin of the cluster of eight events within a $30^\circ$ of the galactic center, apparent in Fig~\ref{hesemap}. Its flux would exceed the point source limits from dedicated searches~\cite{Adrian-Martinez:2014wzf,Aartsen:2014cva}. Correlation of the 3-year event sample with the galactic plane is also not significant: letting its width float freely, the most significant correlation was found for a $7.5^\circ$ width search window, with a post-trials chance probability of 2.8\%.

In this context, one should not forget the strong, and almost certainly unrealistic assumptions underlying the connection between the point and diffuse fluxes. Besides the $E^{-2}$ spectral shape, it is routinely assumed that the flux is time-independent which does not apply to extragalactic sources, and that the sources are not extended which does not apply to some galactic sources. 

The high galactic latitudes of many of the high-energy events do suggest an extragalactic component at some level. We have therefore also searched for clustering of the events in time and investigated correlations with the times of observed GRBs. No statistically significant correlation was found in either search.

Where astronomy is concerned, the highest energy events in Fig.~\ref{diffusenumu} represent a sample of $\sim10$ muon neutrinos with very little atmospheric background. These events can be reconstructed with better than $0.4^\circ$ resolution but were already included in the ongoing IceCube search for point sources which currently shows no evidence for clustering. The search for sources using these events in conjunction with multi-messenger followup observations is interesting and planned for these locations and future neutrino track locations.

Speculations on the origin of the astrophysical neutrinos have been many. Possible source candidates include galaxies with intense star
formation~\cite{Loeb:2006tw,Murase:2013rfa,He:2013cqa,Anchordoqui:2014yva,Chang:2014hua}, cores of active galactic nuclei (AGN)~\cite{Stecker:1991vm,Stecker:2013fxa}, low-luminosity AGN~\cite{Bai:2014kba,Kimura:2014jba}, blazars~\cite{Tavecchio:2014eia,Padovani:2014bha,Dermer:2014vaa},
low-power GRBs~\cite{Waxman:1997ti,Murase:2013ffa,Ando:2005xi}, cannonball GRB~\cite{Dado:2014mea}, intergalactic shocks~\cite{Kashiyama:2014rza}, and
active galaxies embedded in structured regions~\cite{Berezinsky:1996wx,Murase:2013rfa}.
galactic scenarios are also viable and include the diffuse neutrino emission of galactic
cosmic rays~\cite{Ahlers:2013xia,Kachelriess:2014oma}, the joint emission of galactic
PeVatrons \cite{Fox:2013oza,Gonzalez-Garcia:2013iha} or microquasars~\cite{Anchordoqui:2014rca}, and extended galactic structures like
the Fermi Bubbles \cite{Razzaque:2013uoa,Ahlers:2013xia,Lunardini:2013gva} or the galactic
halo~\cite{Taylor:2014hya}. A possible association with the sub-TeV diffuse galactic $\gamma$-ray
emission \cite{Neronov:2013lza} and constraints from the non-observation from diffuse galactic
PeV $\gamma$-rays \cite{Gupta:2013xfa,Ahlers:2013xia}, have also been investigated.
More radical suggestions include PeV dark matter decay scenarios~\cite{Feldstein:2013kka,Esmaili:2013gha,Bai:2013nga,Cherry:2014xra}.

In summary, the presence of events at large galactic latitudes and the absence of significant event clusters suggests extragalactic sources although a galactic component cannot be excluded. Clearly, larger event samples will provide an unprecedented view  into the non-thermal universe.
The resolution of the origin of the observed astrophysical neutrino flux represents a great potential for discovery, but as with any new window on the universe,  we have to be ready for surprises when we will be able to view the sky with more powerful instruments.

\section{\icx{}: From the discovery of cosmic neutrinos to neutrino astronomy\label{sec:astro}}

\subsection{Resolving the sources of astrophysical neutrinos\label{sec:astro:resolving_sources}}

Given that the level of the neutrino flux observed is quite high compared to the diffuse high-energy gamma-ray flux, identifying the sources in multi-messenger and stacking analyses using astronomical catalogues is undoubtedly promising.   Yet the search for point sources using neutrinos only has resulted in upper limits on the flux of individual galactic and extragalactic source candidates~\cite{Abbasi:2011ai,AdrianMartinez:2012rp,Abbasi:2012zw,Aartsen:2013uuv,Adrian-Martinez:2014wzf,Adrian-Martinez:2013dsk}. This may suggest that the observed cosmic neutrinos originate from a number of relatively weak sources. It is indeed important to keep in mind that the interaction rate of a neutrino is so low that it travels unattenuated over cosmic distances through the tenuous matter and radiation backgrounds of the universe. The fact that neutrinos, unlike photons and cosmic rays, have no horizon makes the identification of individual point-sources contributing to the IceCube flux challenging~\cite{Lipari:2006uw,Becker:2006gi,Silvestri:2009xb}. 
IceCube in its present configuration is sensitive to rare transient source classes like GRBs within 5 years of operation via the observation of neutrino multiplets. Identification of time-independent sources is more challenging due to larger backgrounds. We estimate that during the same period IceCube is sensitive to sparse sources such as galaxy clusters by association of events with the closest 100 sources of an astronomical catalogue. A next-generation neutrino observatory with 5 times the point-source sensitivity of IceCube and otherwise similar detector performance would increase the sensitivity to source densities and rates by about two orders of magnitude~\cite{Ahlers:2014ioa}.

Despite the degraded angular resolution and the reduced potential for astronomy, the observation of electron and tau neutrinos to determine the flavor composition of the astrophysical neutrino flux should be a priority.   Additionally, they complement the sky coverage of muon neutrinos at PeV energy, with wide acceptance over the southern sky as illustrated by the HESE analysis.  At high energies the production and decay of unstable nuclei, {\it e.g.}~neutrons with $n\to pe^-\bar\nu_e$, or mesons, {\it e.g.}~$\pi^+\to\mu^+\nu_\mu$ can be the origin of a neutrino flux. Note that the neutrino production from the decay of muons $\mu^+\to e^+\nu_e\bar\nu_\mu$ can be suppressed relative to pion decay if synchrotron losses are important. Hence, the flavor composition is likely energy dependent and can provide insight into the relative energy loss of high energy pions and muons in the magnetic field of cosmic
accelerators~\cite{Kashti:2005qa}.

\begin{figure}
\includegraphics[width=\linewidth]{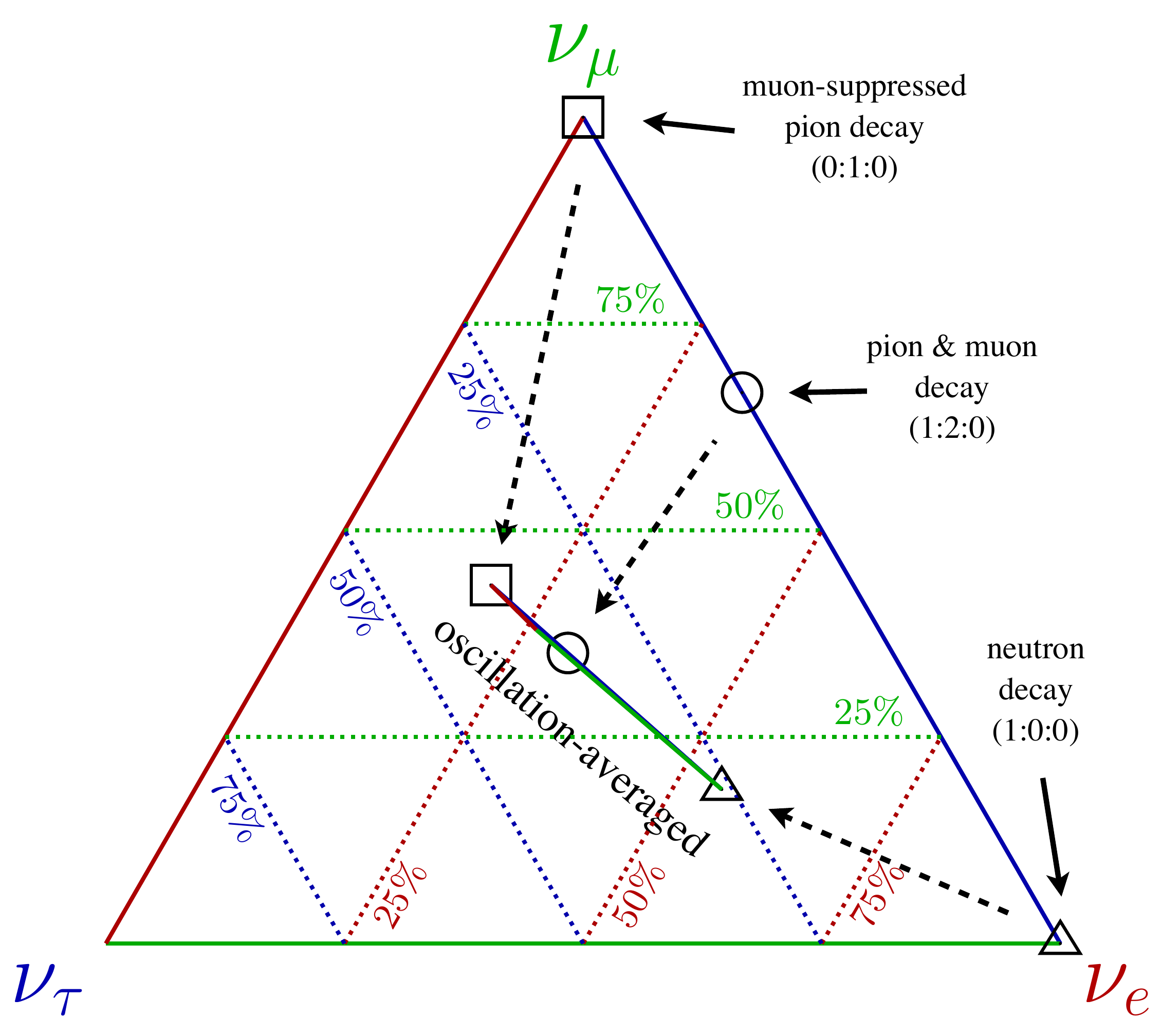}
\caption[]{Neutrino flavor phase space after oscillation. We use the best-fit oscillation parameters
$\sin^2\theta_{12}=0.304$, $\sin^2\theta_{23}=0.577$, $\sin^2\theta_{13}=0.0219$ and $\delta=251^\circ$
following Ref.~\cite{GonzalezGarcia:2012sz} updated after {\it Neutrino 2014}~\cite{nufit}. Each position
in the triangle parametrizes a general initial flavor ratio $(\nu_e:\nu_\mu:\nu_\tau)$. We also indicate specific ratios for neutron decay and pion production. The narrow band in the center is the corresponding observable phase space of oscillation-averaged flavors.}\label{figosc}
\end{figure}

Fig.~\ref{figosc} shows the neutrino flavor phase space $\nu_e$:$\nu_\mu$:$\nu_\tau$ and the expected intrinsic flavor ratio in astrophysical sources from neutron decay (triangle), pion+muon decay (circle) and muon-damped pion decay (square). The observable neutrino flavor ratio is expected to be averaged over many oscillations. This leaves only a very narrow flavor composition range which is shown as the band inset in the center of Fig.~\ref{figosc}~\cite{Barger:2014iua}. 

Flavor identification is less challenging for neutrino energies greater than $1\,\mathrm{PeV}$. The decay length of the $\tau$ produced in charged current (CC) $\nu_\tau$ interactions can be resolved by a neutrino observatory.  Electron anti-neutrinos $\bar\nu_e$ can resonantly interact with in-ice electrons via the Glashow resonance, $\bar\nu_e e^-\to W^-$, at neutrino energies of about $6.3$~PeV. This would be observable as a peak in the cascade spectrum, depending on the relative contribution of $\bar\nu_e$ after oscillation.
In principle, the neutrino-to-anti-neutrino ratios will allow us to answer the basic question of whether the cosmic neutrinos are photo- or hadro-produced in the source~\cite{Anchordoqui:2004eb,Bhattacharya:2011qu,Barger:2014iua}.

\subsection{Neutrinos from the Highest-Energy Cosmic Rays}\label{sec:GZKnus}

The cosmic ray spectrum extends to energies of $10^{20}$~eV. Over Hubble time the highest energy cosmic rays have 
interacted with microwave background photons producing neutrinos. These so-called GZK neutrinos are the decay products of secondary pions, {\it e.g.}, from resonant $\gamma+p \to \Delta^+\to n+\pi^+$ interactions~\cite{Beresinsky:1969qj}. The neutrino flux is calculable because we know the flux of the beam measured by Auger~\cite{Abraham:2010mj,Aab:2013ika} and Telescope Array~\cite{AbuZayyad:2012ru}, the density of the photon target, $410~{\rm cm}^{-3}$, and the $p\gamma$ cross section. In fact, the same interactions that produce the neutrinos limit the propagation of the cosmic rays in the microwave background to a horizon of less than 200\,Mpc. This is the origin of the Greisen-Zatspin-Kuz'min (GZK) suppression~\cite{Greisen:1966jv,Zatsepin:1966jv}.

While calculable, the predicted flux depends on the cosmic evolution of the still-unknown accelerators, and on the maximum energy and composition of the cosmic ray beam. Assuming protons, the rate of GZK neutrinos detected by IceCube is of order one event per year for the most optimistic models.  The IceCube collaboration has performed a series of GZK neutrino searches~\cite{Abbasi:2010ak, Abbasi:2011ji, Aartsen:2013dsm}. The null observation of neutrino induced events above 100~PeV in a 2 year sample of data from the completed detector~\cite{Aartsen:2013dsm} excludes GZK neutrino production by source populations with strong cosmological evolution~\cite{Yoshida:2012gf}, {\it e.g.}~Fanaroff-Riley type II radio galaxies~\cite{Rachen:1992pg}. With the 5 years of data collected so far, IceCube is sensitive at the 68\% C.L. to sources following the Star Formation Rate (SFR). Even the simplest extension of IceCube will result in rates increased by a factor 5 or more, and would predict a guaranteed observation if it were not for one assumption made so far: that the cosmic rays are protons.

Experiments disagree on the composition of UHECRs~\cite{Barcikowski:2013nfa}. In general, a heavier composition tends to lower the predictions of GZK neutrinos~\cite{Hooper:2004jc,Ave:2004uj,Hooper:2006tn,Allard:2006mv,Anchordoqui:2007fi,Aloisio:2009sj,Kotera:2010yn,Decerprit:2011qe,Ahlers:2011sd}. However, even if the contribution of protons to UHECRs is only at the 10\% level
the flux of GZK neutrinos will be observed by a second-generation detector in 5 years for sources with SFR cosmological evolution~\cite{Ahlers:2012rz}. It is interesting that the measurement of the GZK neutrino flux will provide constraints on the chemical composition of the highest energy cosmic rays independent from any conclusions reached from direct measurements by air shower detectors.

Additionally, an extended detector could observe an EeV astrophysical flux at the level of $E^2 \phi_{\nu} \leq 1.0 \times 10^{-9}$~${\rm GeV cm^{-2} s^{-1}sr^{-1}}$ per neutrino flavor, an order of magnitude below the cosmic neutrino flux observed at PeV energy. The complementary information extracted from observation of neutrinos in the PeV and EeV energy ranges should be powerful.

The Askaryan Radio Array (ARA) experiment \cite{Allison:2011wk} is a proposed large scale radio detector at the South Pole aiming to detect the radio emission from neutrino interactions in the 200MHz-800MHz frequency range where the signal is coherently enhanced by the Askaryan effect~\cite{Askaryan1962}. ARA will be sensitive to neutrinos above $\sim$100~PeV. Because of the kilometer--scale attenuation length of radio waves in ice, close to an order of magnitude larger than that of optical light, a radio array can be built in a cost-effective way with spacing of detector stations in shallow ice (200~m depth) of more than 1 kilometer. Since ARA is co-located with IceCube, a combined analysis of ARA and \icx{} creates the opportunity for the simultaneous detection of neutrinos by their optical and radio Cherenkov emission. Hybrid events would result in a very valuable cross--calibration of these different detection methods.

Undoubtedly, the most spectacular impact of the observation of EeV neutrinos would be the measurement of the neutrino cross section testing the electroweak model at a center-of-mass energy of 100 TeV, out of reach of accelerators. A sufficient number of events in either optical or radio detectors would allow a separation of the GZK flux and the neutrino cross section exploiting the measured zenith angle distribution.

\subsection{Multi-messenger studies of transient phenomena}

Multi-messenger astronomy, the confrontation of observations of cosmic rays, neutrinos, photons of all wavelengths, and, in the near future, gravitational waves will represent a powerful opportunity to decipher the physical processes that govern the non-thermal universe. The questions that can be addressed by multi-messenger astronomy are far-reaching: they include the origin of the ultra-high-energy cosmic rays~\cite{Roulet:2012rv,Ahlers:2011sd}, the identification of the processes leading to the highest-energy gammas and neutrinos~\cite{Murase:2013rfa}, the particle acceleration mechanisms~\cite{Dermer:2009her} and the still enigmatic mechanisms powering gamma-ray bursts (GRB) \cite{Razzaque:2004yv,Ando:2005ka}. Especially the multiwavelength observation of the next galactic supernova explosion with the unprecedented neutrino statistics of \icx{} represents an extraordinary opportunity to decipher the complex astrophysics and neutrino physics of the collapse of stars.

The next decade will witness a giant leap forward for multi-messenger astronomy. Large facilities will become operational that cover the electromagnetic spectrum from radio to gamma-rays with LOFAR~\cite{vanHaarlem:2013dsa} and the {\it Square Kilometer Array} (SKA,\cite{Carilli:2004}), the {\it Zwicky Transient Facility} (ZTF,~\cite{ZTF_Proceedings}) and the {\it Large Synoptic Survey Telescope} (LSST,~\cite{Abell:2009aa}) that will allow fast surveys to detect transients from supernovae and GRB afterglows. The {\it High Altitude Water Cherenkov Detector} (HAWC,~\cite{Abeysekara:2013tza}) has started taking data and the {\it Cherenkov Telescope Array} (CTA,~\cite{Acharya:2013sxa}) will be a major step forward in gamma-ray astronomy with order-of-magnitude sensitivity improvements over present generation instruments. Second-generation gravitational wave detectors such as the {\it Advanced Laser Interferometer Gravitational Wave Observatory} (LIGO,~\cite{Abbott:2007kv}) and {\it Advenced Virgo} (\cite{aVirgo}) are likely to directly observe gravitational waves for the first time thus greatly enhancing the potential of ongoing campaigns that already combine gravitational and neutrino signals in order to produce sub-threshold sensitivity~\cite{Aartsen:2014mfp}.

With its improved sensitivity, \icx{} will be a unique instrument to complement these facilities. Neutrinos play a central role in multi-messenger astronomy, as they are an unambiguous signature for the acceleration and interaction of protons and nuclei. They provide an unobstructed and intact view of the sources because they escape extremely dense environments that are opaque to gamma rays, and are unaffected by the intergalactic medium between source and observer. IceCube has a long-standing coordinated observation program with the current generation of Cherenkov telescopes VERITAS and MAGIC~\cite{ICRC:2013ps}. They have performed  follow up searches for interesting neutrino events. The VERITAS telescopes have been pointed in the directions of some of IceCube's highest energy muon neutrinos. Given the likely variability of extragalactic sources, we are enhancing this program to achieve observations within minutes of an interesting cosmic neutrino event.

Neutrino telescopes provide an uninterrupted view over the complete sky with an on-time of more than 99\%. The search for neutrino emission from GRBs in coincidence with past astronomical alerts provides an example of the power of this technique. IceCube data have challenged the idea that GRBs are the origin of the extragalactic cosmic rays~\cite{Milgrom:1995um,Waxman:1995vg,Vietri:1995hs} and clearly ruled out models where neutrons escaping the fireball decay, producing the observed neutrino spectrum~\cite{Abbasi:2012zw}.  \icx{} will be able to make a robust statement on the connection between GRBs and UHECRs.

Some GRBs have been classified as extreme supernova explosions~\cite{Woosley:1993wj,Paczynski:1997yg,Woosley:1998hk}, although no evidence for jet formation is found in regular supernovae.  The missing link between these two phenomena might be a class of supernovae that develops only mildly relativistic jets that fail to penetrate the outer layers of the progenitor star. To test this scenario the directions of interesting muon neutrino events in IceCube are already sent to a variety of optical and X-ray telescopes which search for evidence of supernovae at these positions~\cite{Abbasi:2011ss}. \icx{} will allow us to extend this program in the 2020s, overlapping with a new generation of very powerful wide-field optical telescopes
like ZTF and LSST that survey large parts of the sky every few days. Even if only 1\% of supernovae develop jets, \icx{} will be able to detect several neutrinos per year correlated to such events~\cite{Ando:2005ka}, hence shedding light onto the supernova-GRB connection.

IceCube currently provides the world's most precise opportunity for measuring the time evolution of a galactic supernova explosion~\cite{Abbasi:2011ss}. The signal is predominantly produced by Cherenkov photons emitted from neutrino induced positrons in the volume of ice immediately surrounding each DOM.  The subsequent simultaneous rate increase in all IceCube sensors is statistically extracted above a floor of dark noise that is of the order of only 500 Hz per DOM. IceCube will observe one million neutrinos from a supernova near the galactic center; its sensitivity matches that of a background-free megaton-scale supernova search experiment. The sensitivity decreases to 20 standard deviations at the galactic edge (30 kpc) and 6 standard deviations at the Large Magellanic Cloud (50 kpc). Although this method of detection prevents the details of individual neutrino interactions, {\it i.e.}~the neutrino type, energy and direction, from being identified, we have shown~\cite{Aartsen:2013nla}, that the average neutrino energy can be traced to better than 10\% accuracy for distances less than 10 kpc.

The IceCube high energy extension would more than double the number of IceCube photosensors. Additionally, PMTs with 30\% higher quantum efficiency than that of regular IceCube PMTs (albeit with 25\% increased dark noise) would be deployed. The sensitivity for a supernova collapse in the region of the Magellanic Clouds would increase to $\sim10$ standard deviations, helping to resolve more details of the supernova neutrino light curve.  Additionally, the more than $20$ years of planned operation would significantly increase the chance to observe a galactic supernova.

\subsection{\icx{}: A Tool for Physics}

IceCube has produced world-best limits on the cross section for spin-dependent dark matter particles interacting with ordinary matter for the dark matter mass range of 50 GeV to 5 TeV. They are derived from the failure to observe the production of neutrinos from dark matter particles that annihilate when gravitationally trapped by the sun~\cite{Aartsen:2012kia}. Having no alternative astrophysics explanation, such an observation would represent incontrovertible evidence for dark matter. Plans for the PINGU low-energy extension of IceCube will extend such searches to masses of a few GeV~\cite{Aartsen:2014oha}. Indirect neutrino searches have also used the Earth, the galactic halo~\cite{Abbasi:2011eq,Aartsen:2014hva} and galaxy clusters as targets. As for direct searches, even incremental increases in sensitivity are important given the fundamental physics addressed. This applies to other topics ranging from new particle searches~\cite{Aartsen:2014awd} to the search for the signatures of TeV gravity models and extra dimensions\cite{Kowalski:2002gb,AlvarezMuniz:2002ga,Anchordoqui:2006fn,Illana:2006mv}. Physics beyond the standard model can also be probed by our measurements of atmospheric neutrino interactions in a new energy regime\cite{Illana:2007ra,Illana:2005pu,Albuquerque:2006am,Reynoso:2007zb}, including the EeV energy range of GZK neutrinos previously discussed. 

The surface array IceTop~\cite{Aartsen:2013wda} has proved to be a valuable component of IceCube at a modest marginal cost of approximately 5\%.  Accordingly, for the high-energy array we include in the baseline planning a single surface detector near the top of each  deployed string. With a spacing of 250 m, such a surface array will provide a high-resolution measurement of the primary spectrum from 10 PeV to above one EeV. Most importantly, with the larger aspect ratio of \icx{} high-energy array the acceptance for coincident events seen by both the surface array and the deep array increases
by a factor of 40, from $0.26$~km$^2$sr to $\sim 10$~km$^2$sr. The ratio of the signal of $\sim$TeV muons in the deep detector to the size of the surface shower will allow an unprecedented measurement of the evolution of the primary composition in the region where a transition from galactic to extragalactic cosmic rays is expected.

A surface array also acts as a partial veto for cosmic-ray and atmospheric neutrino backgrounds to high-energy neutrinos in the deep detector. While IceTop provides a veto that covers only about 3\% of the southern sky, with the larger aspect ratio of the high-energy array,  approximately 20\% of the southern sky will have surface coverage. Strategies are under study to extend the surface array beyond the footprint of the array. A larger surface array would enable extended searches for $>$PeV gamma-rays\cite{Aartsen:2012gka} and larger acceptance for  measurements of cosmic-ray anisotropy~\cite{Aartsen:2012ma}.

\section{Design of \icxlong{}\label{sec:design}}

The IceCube Neutrino Observatory builds on the experience with
AMANDA \cite{Halzen:1998bp} and has substantially improved the understanding of the
Antarctic ice as a detection medium for neutrinos.
This understanding has reached a maturity which allows us to perform
searches which had not been anticipated in the initial design.
These searches include the precision measurement of neutrino oscillations,
the reconstruction of cascade-like events with about
$10^\circ$ directional resolution and few percent energy resolution,
and full sky searches for cosmic neutrino point sources
with an angular resolution below $0.5^\circ$, all performed with
a sensitivity that surpasses the original design by a factor 2 or more.
These successes developed the expertise and provided the experience
to design the next generation instrument with a substantially
increased detector volume.

As part of building and calibrating the IceCube detector, the optical properties of natural ice over large distances
were carefully measured, with the surprising discovery that the absorption length of the Cherenkov light to which the DOMs
are sensitive exceeds $100\,\mathrm{m} - 200\,\mathrm{m}$, depending on depth. 
Therefore, it is possible to instrument a significantly larger volume of ice with lower string densities than used in 
IceCube.

The larger spacings do of course result in a higher energy threshold.
While the 100,000 or so atmospheric neutrinos that IceCube collects above a threshold of 100\,GeV every year
are useful for calibration and neutrino physics studies, they also represent a background for isolating the cosmic component
of the flux. The peak sensitivity to an $E^{-2}$ spectrum is reached at 40\,TeV \cite{Halzen:2008zj}.
While the IceCube detector has to be efficient below that energy, a detector with a higher energy 
threshold can be considered without loss of astrophysical neutrino signal.

Design work is now underway on this next generation neutrino observatory, \icx{}, with the goal to instrument
a volume of $10{\rm km}^3$, and deliver a substantial increase in sensitivity to astrophysical neutrinos
of all flavors.  \icx{} would build upon the existing IceCube detector infrastructure, and would take
advantage of the very long absorption lengths found in the glacial ice near the IceCube detector to 
add additional instrumentation with significantly larger string separation distances.   Building this larger
instrument would be achieved with a comparable number of strings used in the existing IceCube detector,
and would target neutrino energies above $\sim50$ TeV with high efficiency.  

While detailed studies toward a design are ongoing, several design considerations are clear and understood.  
A detector sensitive to high-energy astrophysical muon, electron and tau neutrino flavors requires the correct combination of 
instrumented volume and projected surface area in all directions.   This optimization is done by selecting larger string spacings,
which increases the energy threshold as the instrumented volume increases, and geometrical arrangement of the deployed strings,
which can trade detector instrumented volume for detector cross-sectional area.  Detection of neutrino-induced muon track events
will scale with the detector cross sectional area, while the neutrino shower events produced by electron and tau flavors, and neutral current
interactions will scale with instrumented volume.
These studies will culminate in a string arrangement and detector design that delivers
a next generation instrument with optimal sensitivity to all flavors of neutrinos.

The most important aspect of a larger instrument in the glacial ice are the optical properties of the ice,
where the measured absorption length of the Cherenkov light to which the DOMs
are sensitive exceeds $100\,\mathrm{m}$. In fact, in the lower half of the detector it exceeds $200\,\mathrm{m}$. 
Although the optical properties vary with the layered structure of the ice, the average absorption and scattering lengths dictate
the distance by which one can space strings of sensors without impacting the uniform response of the
detector. Early studies indicate that spacings of $\sim240\,-\,300\,\mathrm{m}$ are acceptable
while maintaining high efficiency to astrophysical neutrinos.

\begin{figure}
\includegraphics[width=0.5\textwidth]{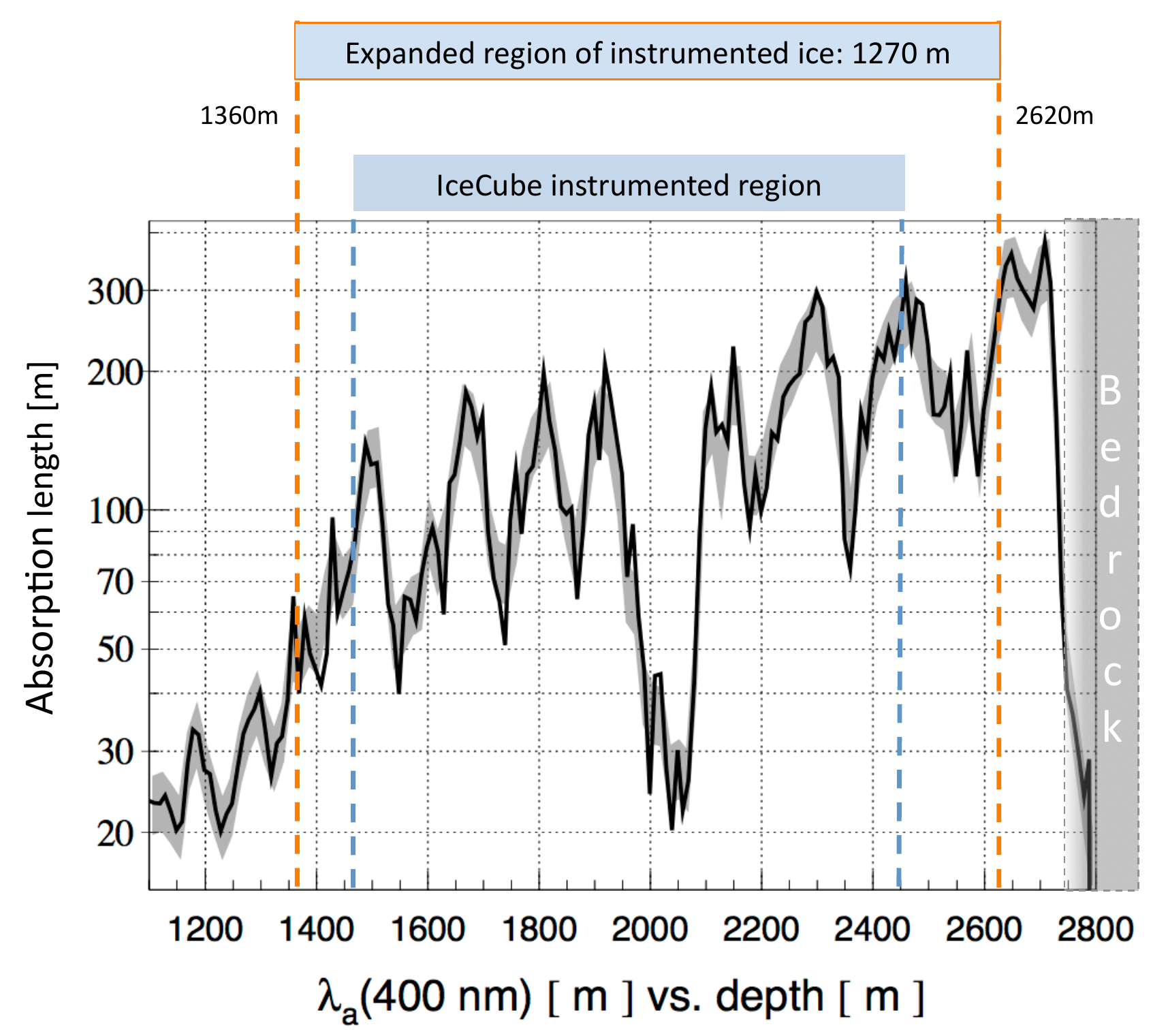}
\caption{Absorption length in the glacial ice versus depth. Note the layer of high
dust concentration starting at about 2000m depth. The ice above and below that layer
is very clear. The current instrumented depth range used in IceCube and an extended
string length, adding about $240\,\mathrm{m}$ to each string are indicated. Note that not all
simulations shown in this report have been performed with the extended string length.}
\label{fig:design:absorptivity}
\end{figure}

The optical properties of the glacial ice prevent us from using optical modules at depths much shallower than the current instrumented 
range used by IceCube (with instrumented depths between ~1450m and 2450m).  
From the depth dependence of the absorptivity of the Antarctic ice (shown in Fig. \ref{fig:design:absorptivity}), 
we will be able to extend the strings by 75\,m at the top and 175\,m at the bottom. This string length extension leads to a 25\% increase
in the geometric area for horizontal track events and therefore a 25\% increase
in effective area for such events. 

Another key for the successful scientific
operation of IceCube has been the excellent reliability of the
Digital Optical Modules (DOMs) \cite{Abbasi:2010vc},
drilling and DOM deployment operations \cite{Abbasi:2010vc} and data systems \cite{Abbasi:2008aa},
which have resulted in a stable data taking at more than 99\% uptime.
We will base our studies conservatively on these successful
technologies, only slightly modified to newer components and higher efficiencies.
The main method to achieve a significant enlargement
of the next generation instrument are 
an increased horizontal spacing between strings and an increased instrumented length
for each string, either by increasing the vertical distance between DOMs or
adding additional DOMs to each string.

\begin{figure*}
 \subfloat[$240\,\mathrm{m}$ string spacing (``benchmark'')]{\label{fig:design:geometries:sunflower120_240}%
  \includegraphics[width=0.45\textwidth]{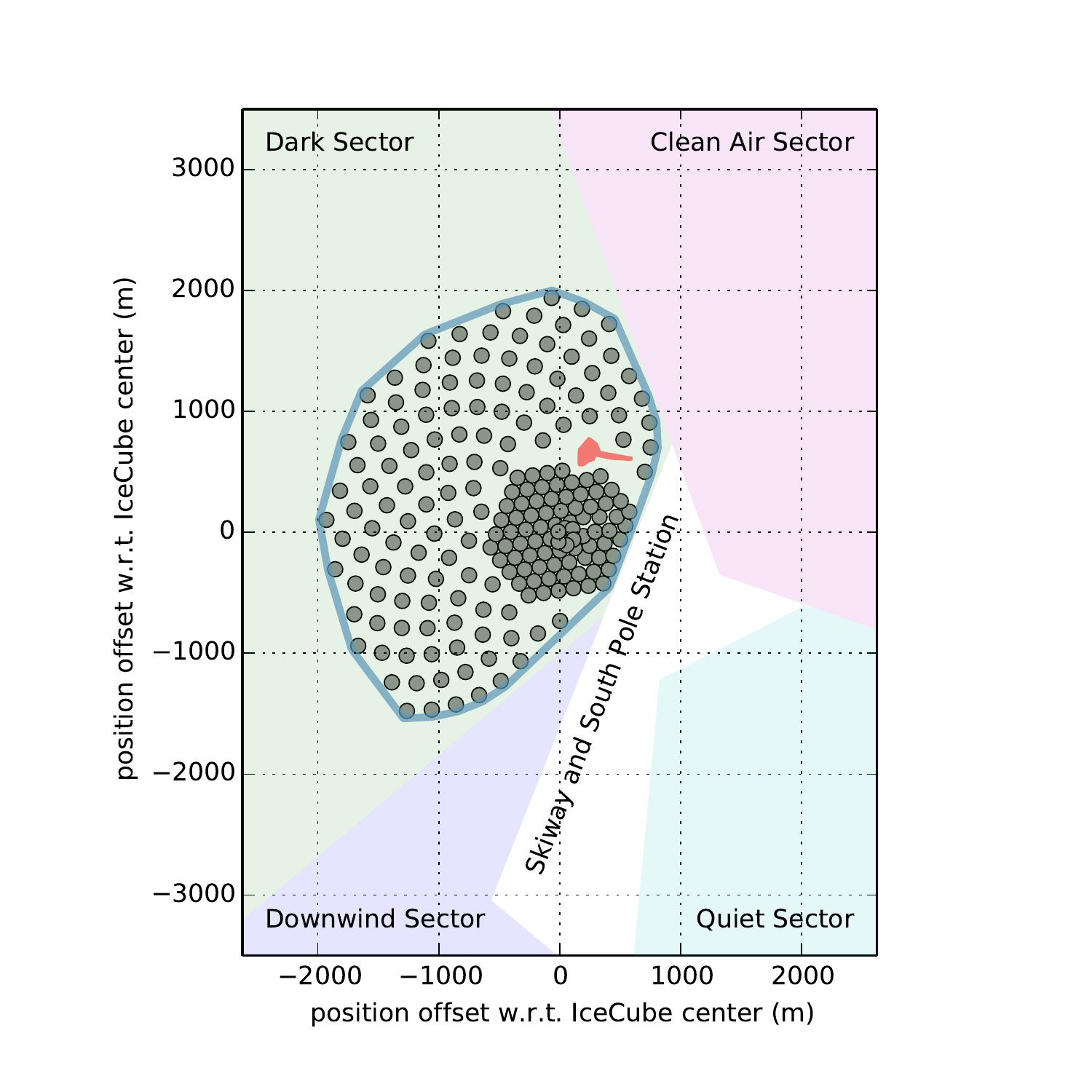}%
 }%
 \subfloat[$300\,\mathrm{m}$ string spacing]{\label{fig:design:geometries:sunflower120_300}%
  \includegraphics[width=0.45\textwidth]{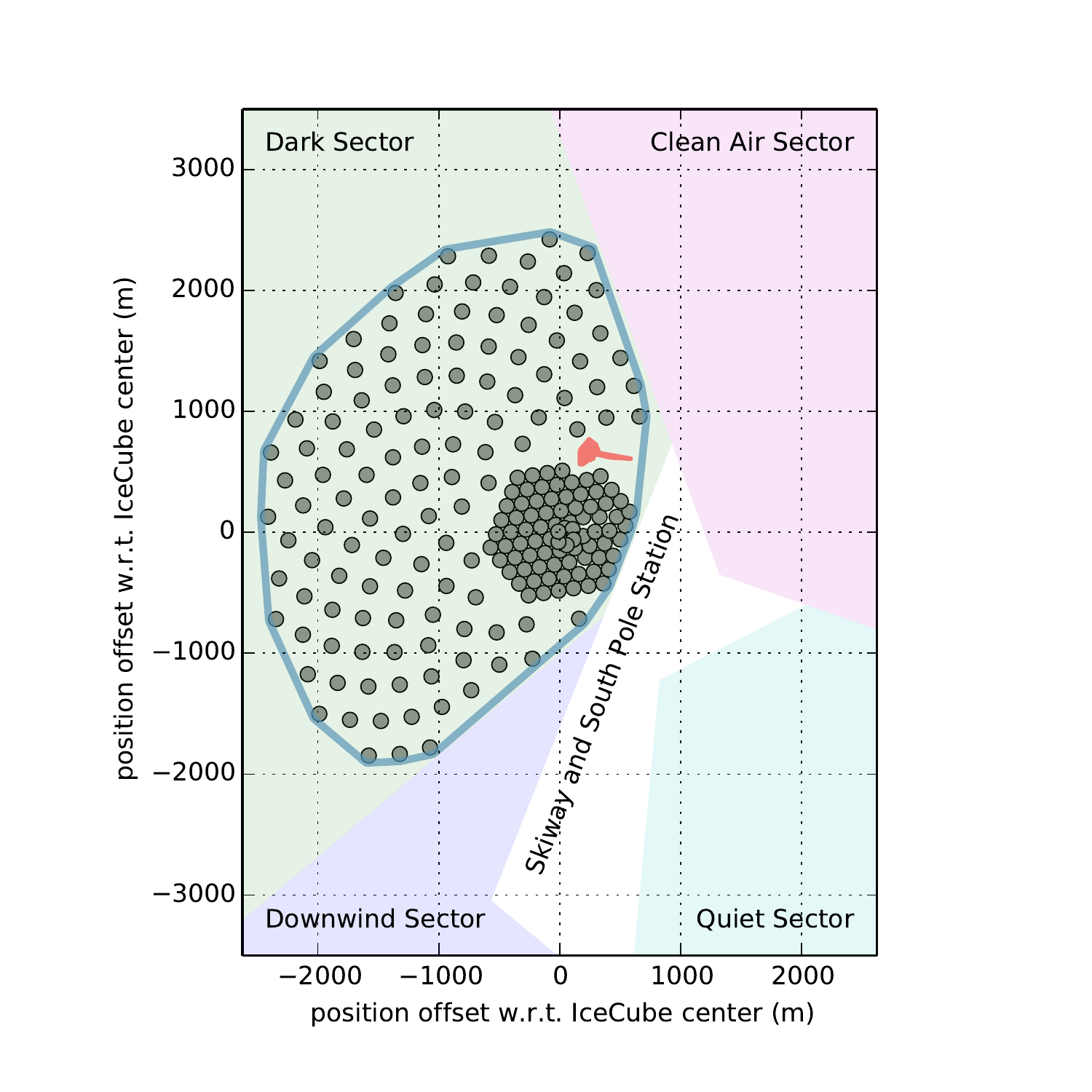}%
 }
  \caption{The benchmark detector string layout studied in this paper is shown
  in sub-figure (a) and extends IceCube by 120 strings constrained to the South Pole
  ``Dark Sector'' (shaded in light green).
  The string-to-string distance for the detector shown is $\sim 240\,\mathrm{m}$.
  This is not a final design and represents only
  one of the layouts under consideration. It is shown here as an example for a possible
  upgraded detector. Other geometries and string spacings are under consideration.
  A possible variant of the benchmark detector is shown in sub-figure (b) and
  differs in string spacing ($\sim 300\,\mathrm{m}$). This and other variants are under active study.}
 \label{fig:design:geometries}
\end{figure*}

To begin investigating the performance of possible future detector designs,
we studied a benchmark detector in simulation:
a compact detector with string placements excluding parts of the South Pole area
currently not easily available for drilling as shown in Fig.~\ref{fig:design:geometries}.
As a baseline spacing for new strings,
we use a distance of approximately $240\,\mathrm{m}$, but other distances up
to $360\,\mathrm{m}$ have also been considered. 
The geometrical placement of strings is still being optimized and different
options are being explored to optimize the physics potential for all neutrino flavors
and event channels, such as incoming tracks and events with contained vertices.

\begin{figure}
\includegraphics[width=\linewidth]{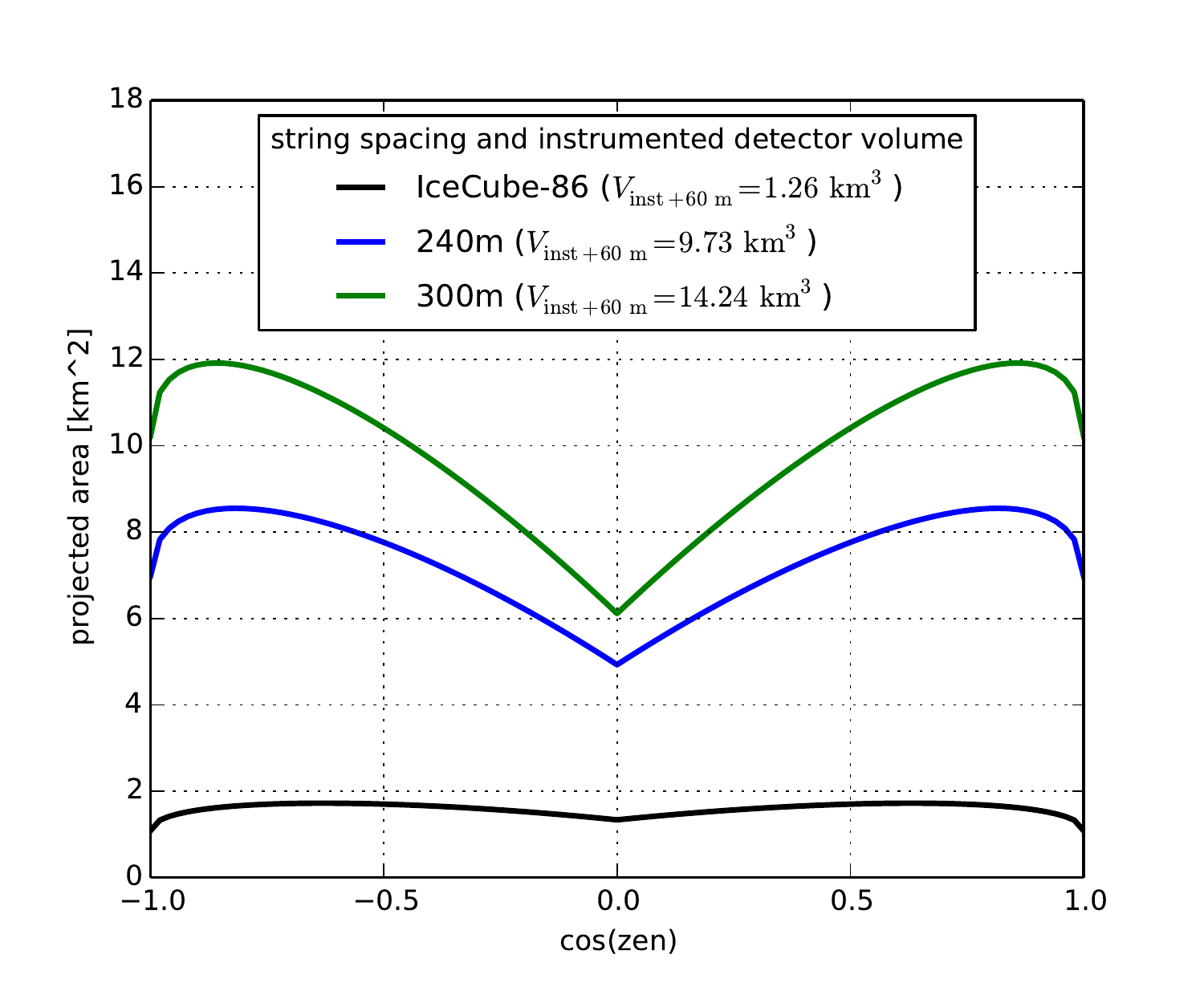}
\caption{Instrumented area as a function of zenith of IceCube, the benchmark detector ($240\,\mathrm{m}$ string spacing)
and its variant with a string spacing of $300\,\mathrm{m}$.
To determine the 
geometric area, the detector volume containing all optical modules has been extended by $60\,\mathrm{m}$ on all sides.
Note the difference in area when viewed from the side ($\cos(\mathrm{zen})=0$)
compared to the top/bottom area.
The figure legend also shows the instrumented volume of each detector (again with an
extension of $60\,\mathrm{m}$ at the sides, top and bottom to account for events outside
of the detector that can still be recorded.)
}
\label{fig:design:geoarea}
\end{figure}

The geometry of this benchmark detector consists of strings on a non-regular grid avoiding symmetries that deteriorate acceptance and resolution for muon tracks.
The benchmark geometry is compared to the IceCube detector in its completed 86 string configuration
and is used to scale sensitivities to a $10\,\mathrm{km}^3$ instrument.
The projected areas of two of the considered geometries are compared to IceCube-86 in Figure~\ref{fig:design:geoarea}.
As the detector volume grows in these geometries, the exposed area increases
and reaches up to $\sim 10\,\mathrm{km}^2$ area, substantially larger than the IceCube
area.

\subsection{Neutrino sensitivity considerations}

Several factors determine the sensitivity to astrophysical neutrinos for each detection channel and neutrino energy range.
These factors include detector volume, projected cross sectional area, energy and angular resolution,  and
background event rates.  The overall sensitivities of each potential detector design need to be
evaluated.  To date, these are done by scaling existing IceCube analysis to the larger
instruments without optimization to the larger detector, which can potentially underestimate the gains in sensitivity of
detector designs.  Additionally, novel background rejection techniques such as a surface
atmospheric event veto can greatly change the overall sensitivity of
the instrument, and are not applied in these studies.

For a traditional point source search, which relies on muon tracks produced 
by charged current interactions of muon neutrinos in or near the instrumented 
volume, the sensitivity increases with the projected cross sectional area relative
to source direction.  At the energies of interest for astrophysical neutrino searches,
these muons have ranges that routinely exceed 10 km, greatly extending the sensitivity
of these searches.

Search sensitivities of the detector will scale approximately with the
square-root of the increase in cross sectional area and linearly with the angular resolution.
For a large portion of the sky, an increase in angular resolution is expected
from the longer track lengths observed in the larger instrument.
The angular resolution can further be improved by using better, more computationally intensive reconstruction methods 
that use a more detailed model of the ice properties in the instrumented volume. 
Combined with an increase in effective area of a factor of $\sim 5$ (assuming optimized
string distances and extended string instrumentation depths), we aim for an increase
in sensitivity of a factor of $\sim 5$ and beyond.

A search for electron or tau neutrino interactions depends on observation of an 
electromagnetic cascade resulting from the interaction of the neutrino
with nucleons inside the instrumented volume.  As the secondary particles
have very short track lengths compared to the instrumentation spacing, these events
appear as nearly spherical light depositions in the detector. 

These neutrinos generally
interact with nucleons in the ice via deep-inelastic scattering (DIS) processes, but
electron anti-neutrinos with an energy of $E_{\nu}\sim6.3\,\textnormal{PeV}$ 
have an enhanced probability to scatter off atomic electrons in the ice by forming
an on-shell $W^{-}$-boson, the so-called Glashow resonance (GR).~\cite{Glashow:1960zz}
The resonance would be observable mostly as a peak in the cascade 
energy spectrum.  

The neutrino energy is determined from the ``visible'' energy $E_{vis}$,
defined as the energy deposit of a purely electromagnetic cascade that
produces the observed Cherenkov light.  The GR events exceed the continuum of DIS induced cascades 
around $E_\nu=6.3\,\textnormal{PeV}$.
An overview of the expected number of DIS and GR contained events for benchmark detector configurations and
different neutrino fluxes is given
in Table~\ref{tab:GR-rates}.
These expectations are lower in the case of pure p$\gamma$ sources
because of the suppressed electron anti-neutrino contribution~\cite{Bhattacharya:2011qu}.
As an example, in IceCube-86 and assuming an $E^{-2}$ spectrum at the measured neutrino flux level, we expect to observe $0.9$ GR contained cascades induced by electron anti-neutrinos with
energies between $5\,\textnormal{PeV}$ and $7\,\textnormal{PeV}$ per year. This is for an assumed
neutrino to anti-neutrino ratio of $(\nu_e:\bar{\nu}_e)=(1:1)$, which is a generic prediction
for pure pp sources~\cite{Aartsen:2013jla}.

\begin{table*}[ht!]
	\centering
\begin{tabular}{c c c c c c c c c}
	\hline\hline
		$\Phi_{\nu_e}$   &  \,\,  interaction \,\, &  \multicolumn{2}{c}{ pp  source}  &   \\ \cline{3-5} \
		[$\, \textnormal {GeV}^{-1} \textnormal{cm}^{-2} \textnormal{s}^{-1}  \textnormal{sr}^{-1}$]  & type & \,\,  IC-86 \,\, &  \,\, 240m & 360m\,\,   \\
	\hline
		\begin{tabular}{r}
	       $1.0 \times 10^{-18}(E/100\,\textnormal{TeV})^{-2.0}$  \\  \\ $1.5 \times 10^{-18}(E/100\,\textnormal{TeV})^{-2.3}$   \\  \\  $2.4 \times 10^{-18}(E/100\,\textnormal{TeV})^{-2.7}$   \\  \\
		\end{tabular} &
		\begin{tabular}{r}
			 GR \\  DIS  \\  GR  \\  DIS   \\  GR  \\  DIS  
		\end{tabular} &
		\begin{tabular}{r}
			0.88 \\  0.09  \\  0.38 \\ 0.04   \\  0.12 \\ 0.01
		\end{tabular} &
		\begin{tabular}{r}
			7.2 \\  0.8 \\  3.1 \\ 0.3   \\  0.9 \\ 0.1
		\end{tabular} &
		\begin{tabular}{r}
			  16 \\  1.6 \\  6.8 \\ 0.7   \\  2.1 \\ 0.2
		\end{tabular} &
	\end{tabular} 
	\caption{Expected number of contained neutrino-induced cascades per year with $5\,\textnormal{PeV} < E_{vis} < 7\,\textnormal{PeV}$ 
	in IceCube in its current 86-string configuration and in an extended detector with a string spacing of 240 m (360 m shown for comparison) 
	assuming a source dominated by p-p interactions.  For every event Cherenkov light is required to be detected by optical modules on at least 3 strings.}
\label{tab:GR-rates}
\end{table*}

\begin{figure}
  \centering
 \includegraphics[width=\linewidth]{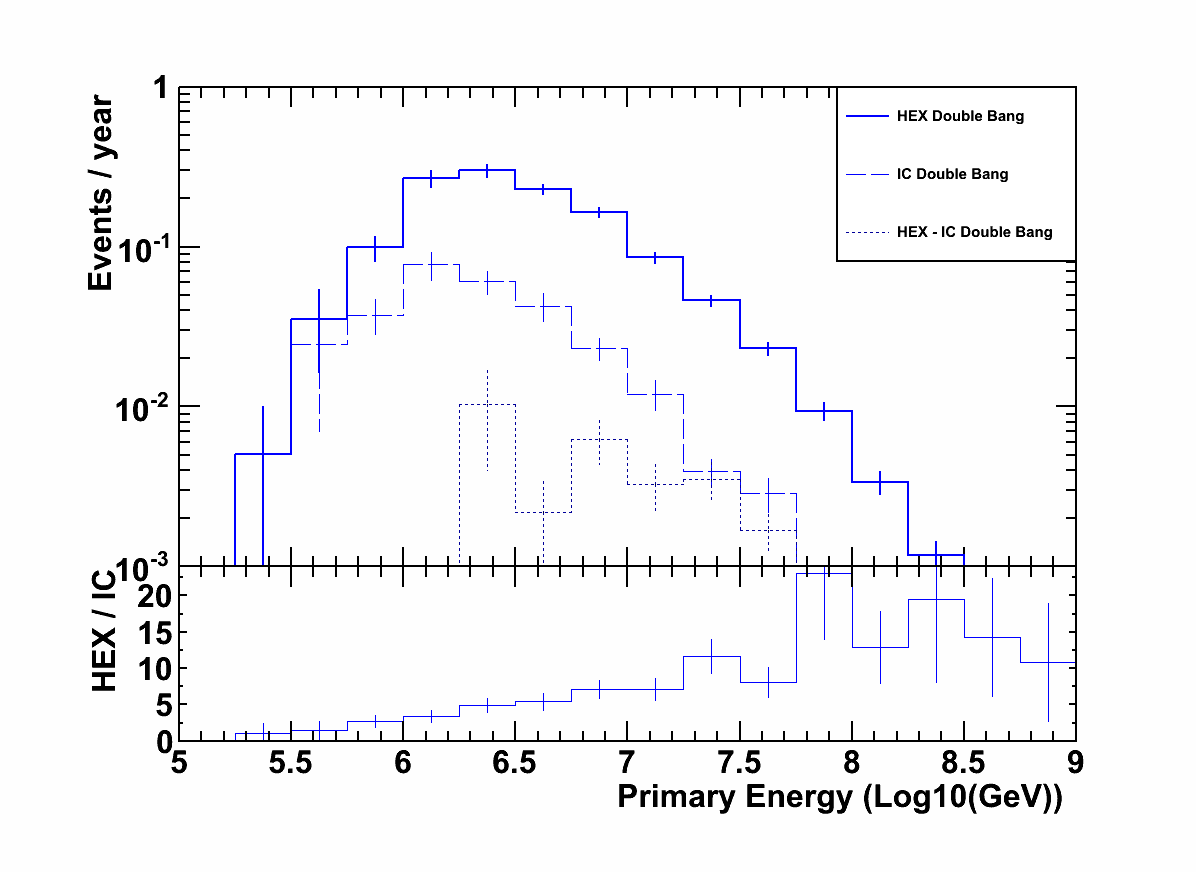}
  \caption{
    Tau Neutrino induced double bang rate per year as a function
    of neutrino energy in IceCube (``IC'', dashed),
    the benchmark extended volume with 240\,m string spacing (``HEX'', solid)
    and in both volumes (``HEX+IC'', dotted).}
  \label{fig:tausense}
 \end{figure}

A signature of a tau neutrino is a cascade from the tau neutrino
charged current (CC) interaction, followed by a second cascade (hadronic or
electromagnetic) from the tau decay. If the cascades are well separated, the signature is
called a double bang~\cite{Learned:1994wg}.
Figure~\ref{fig:tausense} shows the tau neutrino sensitivity of the benchmark design, assuming a neutrino flux
as measured by the 3 year HESE search~\cite{Aartsen:2014gkd} and equal oscillation into all three
flavors at the detector. An extended geometry will yield
a factor of 10 increase in double bang tau
neutrino event rates at PeV energies compared to IceCube.

\icx{} would also be able to perform searches for GZK neutrino (see section~\ref{sec:GZKnus})
by looking for neutrino events with energies above 100~PeV.  
The benchmark design shows that a factor of $\sim5$
increase in event rate can be achieved by a $10\,\mathrm{km}^3$ detector
when compared to IceCube,
where several models predict a single event per year in IceCube for a
proton-dominated extragalactic cosmic ray flux.

\subsection{Vetoing atmospheric backgrounds}

Most of the preliminary studies mentioned above presume that the southern sky
is inaccessible to the detector when considering incoming tracks due to the overwhelming
background of muons from cosmic ray air showers. This background can be greatly
suppressed by dedicating parts of the in-ice instrumentation in order to tag incoming muon tracks.
This method has been used in a growing number
of analyses, including the first observation of an astrophysical neutrino flux~\cite{Aartsen:2013jdh}.

In addition to in-ice veto strategies, cosmic-ray showers can be directly vetoed on the ice surface. 
A surface veto detector, possibly based on technology similar to IceTop (or a simplified version of it), 
and extended to large areas of several $\mathrm{km}^2$ can be used to detect CR air showers, and 
veto the in-ice muons and neutrinos they produce.  
Such a surface veto is not considered in the scope of the benchmark designs presented here, but preliminary
studies have shown that it has the power to greatly increase the sensitivity to sources in the southern sky.

With the addition of a cost-effective atmospheric veto, all-sky neutrino studies would be possible without having to restrict samples
to smaller detector volumes or neutrino energies above atmospheric muon backgrounds.  As the energy
spectrum in the southern sky is not limited by neutrino absorption in the Earth, the potential
for sensitivity gains is large,  and further studies will consider the inclusion of a surface veto component.

\section{Instrumentation \& Deployment\label{sec:instrument}}
Important considerations in the construction of \icx{} are a robust
design for the instrumentation that is deployed into the ice
and improved performance and reliability of the drilling equipment used to deploy that instrumentation into the ice.  The designs of the digital optical module (DOM) and Enhanced Hot Water Drill (EHWD)
used in the IceCube MREFC construction project have proven to be
cost effective and robust.  With almost 5 years of operation
for the completed IceCube detector (and more than 10 years of
operation for the first deployed DOMs), 98.5\% of the IceCube DOMs
are still operating and collecting high-quality physics data. Less than 1\% were lost during deployment, very few during operation. At the peak of IceCube construction, the EHWD efficiently allowed for the deployment of 20 strings in a single season.

The instrumentation and drilling systems for \icx{} are designed closely following the demonstrated IceCube
technologies, with targeted improvements for overall performance and new capabilities.  While the \icx{} baseline DOM keeps the robust structural elements of the IceCube DOM, we will introduce a modern, more powerful set of electronic components into its design. The EHWD design focuses on a modular system that will operate with higher efficiency and require less maintenance in routine operation.  Based on the successes of the IceCube project, the \icx{} DOM and EHWD systems can be developed with
low levels of cost and schedule risk.  Additionally, given the digital nature of the detector concept,
the \icx{} extension to IceCube will be operated together with IceCube as a single facility, and will be operated without a significant increase in cost.

Both the instrumentation design (DOM, cable and readout) and deployment (EHWD) are shared with the PINGU low-energy array as part of the \icx{} infrastructure, providing significant reductions in cost and enhancing flexibility.

\subsection{Advances in optical module and electronics design}
At the heart of the \icx{} design is an updated DOM.  Each DOM consists of
a photomultiplier tube and electronics for digitization, control
and calibration enclosed in a pressure resistant glass sphere.
In operation, each DOM is an independent light collection unit, responsible
for self-triggering, time synchronization, digitization and transmission of
recorded data to the surface DAQ for detector triggering and readout.

\begin{figure}
\includegraphics[width=\linewidth]{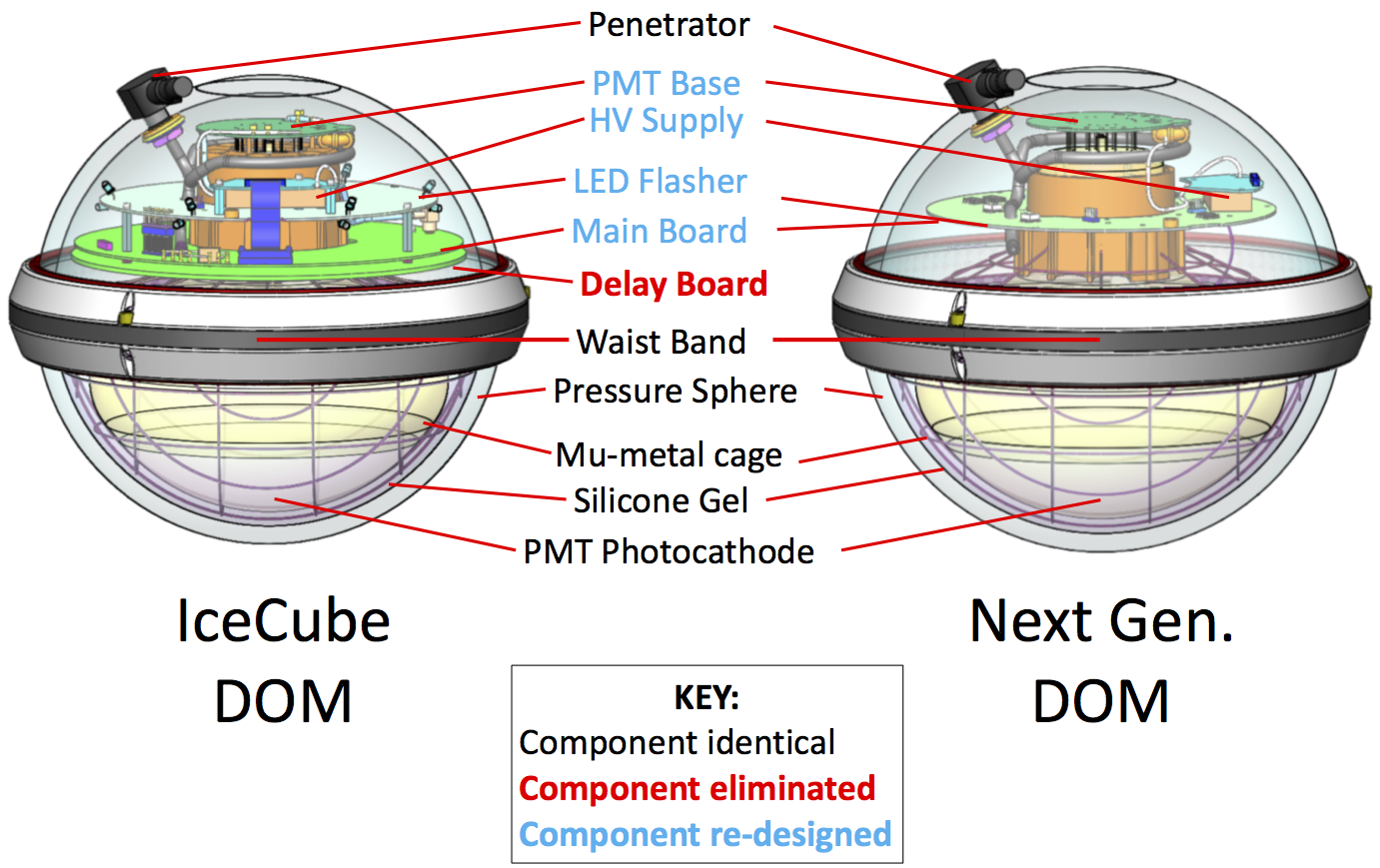}
\caption{Design changes for the digital optical module(DOM) highlighting the
changes from the baseline IceCube DOM design to the next generation DOM
design. Components unchanged are listed in black, redesigned components in
blue and eliminated components in red.  All structural elements remain
unchanged from the successful IceCube design, while the electronics
systems are replaced with updated, equivalent systems.}
\label{fig:NextGenDOM}
\end{figure}

Figure~\ref{fig:NextGenDOM} highlights the design changes for the
next generation DOM.
The design of the \icx{} DOM maintains the same structural elements
used in the IceCube design.  All elements that
are exposed to the intense forces and pressures when the hole refreezes are left unchanged, including the glass pressure sphere,
waist band and cable penetrator.  The \icx{} DOM design uses the high quantum efficiency 10 inch diameter photomultiplier used in the IceCube Deep Core sub-array, as well as the same mu-metal magnetic shield
cage and silicone gel coupling the PMT to the glass pressure
sphere.  By maintaining these proven components from the IceCube
design, many of the environmental and structural design risks are
mitigated.

The high voltage supply, signal digitization, calibration signals
and DOM-to-surface communications will be updated from the original
IceCube DOM designs.  The $\sim$15 year old electronics design of IceCube digitizes the PMT waveforms at 300 MHz using custom ASIC chips (ATWD: Analog Transient Waveform Digitizers,~\cite{Abbasi:2008aa}) under the direction of an FPGA-based system mainboard. Both components, as well as other mainboard components, are no longer available.  The updated \icx{} DOM design will provide
the same functionality by incorporating modern, commercial components, including high-speed ADCs for triggerless
signal digitization, and more powerful FPGA/CPU systems that provide
significantly more functionality with reduced power consumption and an overall simpler design.  Additionally, the surface
electronics supporting the next generation DOM are also redesigned. With the shift to widely used,
commercially available components, the redesign of all electronic systems can
be accomplished with minimal risk, and reduced power consumption and overall cost.  Additionally, the next-generation DOMs will support greater multiplexing for communication with the surface electronics,
allowing for additional cost savings and simplification of the cable
infrastructure of \icx{}.

In parallel, new concepts for optical sensors, employing multiple small PMTs or wavelength-shifting and light-guiding techniques, are 
being studied~\cite{Aartsen:2014oha}. These will collect at least a factor two more Cherenkov photons per module, and while the multi-PMT 
DOM yields intrinsic directional information, the wavelength-shifting module features would produce very low noise levels. PINGU will 
serve as a testing platform to study the performance and reliability of these new concept optical sensors in-situ. In the end, a combination of 
different module types may provide the best sensitivity.

\subsection{Evolution of the hot-water drill}
The EHWD~\cite{IceDrill} was used to create the vertical $\sim$~2.5 km-deep water holes in the Antarctic glacier in which strings of DOMs were deployed at a rate of 20 for a single summer construction season.
The EHWD consists of two primary components, a fern drill used to thaw the compacted snow
near the surface, and a high pressure recirculating hot water drill that delivers 5 MW of thermal
power
 to melt a vertical column of ice to depths greater than 2.5 km into the glacial ice.  While substantial portions of the successful IceCube EHWD remain available for reuse, others have been repurposed. Additionally, the larger spacings between string locations in the \icx{} plan provide logistical challenges for the IceCube EHWD design, where a central
high-pressure hot water plant was erected each season from which hot water was piped
to the 20 string locations. The IceCube EHWD also required a large expert crew for operation and a significant maintenance effort to maintain efficient operation.

The EHWD system considered for \icx{} addresses these issues, providing a more mobile
and efficient high-pressure hot water plant that reuses several existing IceCube EHWD components.
The \icx{} EHWD system is based on a modular design that is sled-mounted to move between groups of strings in a single season.  The high-pressure
hot water plant will consist of several modular water heating and pumping units, each consisting
of a microturbine generator, a heat exchanger for waste heat recapture,  water heaters and a high pressure pump.
Several of these units would work in parallel to provide the needed hot water for deep-ice
drilling.  The proposed system would greatly reduce the complexity of the water heating plant,
enabling the system to operate with a smaller labor force and reduced maintenance.
The remaining components of the  \icx{} EHWD system would be built from IceCube EHWD components,
including the cable and hose reel systems, the drilling and deployment towers and all drilling support systems.  The \icx{} EHWD system will also 
filter and degas the water used in the drilling system to provide better optical properties
in the refrozen ice column.

\section{Conclusions\label{sec:discuss}}

With the detection of an astrophysical neutrino signal, IceCube has detected the ``first light'' in the field of high-energy neutrino astronomy.  
However, detailed spectral studies and searches for specific source locations in this signal remain a challenge with the event sample sizes
available from the current IceCube instrument.  

Design studies for a larger instrument, the \icx{} high-energy array, are well underway. They will result in an instrumented volume approaching 
$10\,\mathrm{km}^3$ and will lead to significantly larger neutrino detection rates, across all neutrino flavor
and detection channels.  With these large astrophysical neutrino samples and the
improved reconstruction and background rejection techniques,  detailed searches for sources and
studies of neutrino spectra are possible, both alone, and in conjunction with the next generation
of electromagnetic and gravitational observatories.   Beyond neutrino astronomy, this instrument will have a 
broad physics program that includes searches for beyond-the-standard-model physics, studies of the properties of
cosmic rays, and searches for neutrinos from nearby supernova.

As previously stated, the path to the next generation neutrino astronomy instrument is clear.   A complete preliminary design for 
\icx{} that combines the robust systems for drilling and detector instrumentation demonstrated with the IceCube detector with an
optimized geometrical sensor arrangement that maximizes sensitivity to astrophysical neutrinos is being finalized now.
Once in operation, the \icx{} high-energy array, as part of the larger \icx{} facility at South Pole will truly be the flagship
for the emerging field of neutrino astronomy.

\begin{acknowledgments}
We acknowledge support from the following agencies:
U.S. National Science Foundation-Division of Polar Programs,
U.S. National Science Foundation-Physics Division,
University of Wisconsin Alumni Research Foundation,
the Grid Laboratory Of Wisconsin (GLOW) grid infrastructure at the University of Wisconsin - Madison, the Open Science Grid (OSG) grid infrastructure;
U.S. Department of Energy, and National Energy Research Scientific Computing Center,
the Louisiana Optical Network Initiative (LONI) grid computing resources;
Natural Sciences and Engineering Research Council of Canada,
WestGrid and Compute/Calcul Canada;
Swedish Research Council,
Swedish Polar Research Secretariat,
Swedish National Infrastructure for Computing (SNIC),
and Knut and Alice Wallenberg Foundation, Sweden;
German Ministry for Education and Research (BMBF),
Deutsche Forschungsgemeinschaft (DFG),
Helmholtz Alliance for Astroparticle Physics (HAP),
Research Department of Plasmas with Complex Interactions (Bochum), Germany;
Fund for Scientific Research (FNRS-FWO),
FWO Odysseus programme,
Flanders Institute to encourage scientific and technological research in industry (IWT),
Belgian Federal Science Policy Office (Belspo);
University of Oxford, United Kingdom;
Marsden Fund, New Zealand;
Australian Research Council;
Japan Society for Promotion of Science (JSPS);
the Swiss National Science Foundation (SNSF), Switzerland;
National Research Foundation of Korea (NRF);
Danish National Research Foundation, Denmark (DNRF).
I.~Bartos, S.~Marka, Z.~Marka, and M.~H.~Shaevitz are thankful for the generous support of Columbia University in the City of New York.
\end{acknowledgments}

\bibliography{paper}

%
%
%
%
%
%
%

\end{document}